\documentclass[a4paper,11pt]{article}
\pdfoutput=1
\usepackage{jheppub}
\usepackage[T1]{fontenc} 
\usepackage{amsmath,amssymb,graphicx,bbm,mathrsfs,nicefrac}
\usepackage{tikz}
\usepackage{amsmath}
\usepackage{amssymb,wasysym, marvosym}
\usepackage{graphicx,textcomp,float,gensymb,wrapfig, enumitem,comment,dsfont,framed,slashed,appendix,wrapfig,xcolor}
\usepackage{ bbold }
\usepackage{bm}
\usepackage{filecontents}
\usepackage{mathtools}
\usepackage{braket} 
\usepackage{hhline}
\usepackage{ mathrsfs }
\usepackage{rotating}
\usepackage[small]{caption}
\usepackage{subcaption}
\usepackage{etoolbox,hyperref,makecell}
\usepackage{feynmp}
\usepackage{accents}
\usetikzlibrary{arrows.meta}
\usepackage{empheq}
\usepackage{filecontents}

\usepackage{multirow}

\usepackage{braket} 

\patchcmd{\abstract}{\null\vfil}{}{}{}
\newcommand{\g}{(g-2)_\mu}

 \newcommand{\brac}[2]{ \left( \frac{#1}{#2} \right) }

\usepackage{pifont}

   \DeclareMathOperator{\gev}{GeV}

\newcommand{\beq}{\begin{equation}} \newcommand{\eeq}{\end{equation}}
\newcommand{\bea}{\begin{eqnarray}} \newcommand{\eea}{\end{eqnarray}}

\newcommand{\fref}[1]{Figure~\ref{#1}}
\newcommand{\eref}[1]{Eq.~(\ref{#1})}
\newcommand{\sref}[1]{Section~\ref{#1}}
\newcommand{\aref}[1]{Appendix~\ref{#1}}

\newcommand{\be}{\begin{eqnarray}} \newcommand{\ee}{\end{eqnarray}}
\newcommand{\Eq}[1]{Eq.~(\ref{#1})}

\usepackage{hhline}
\usepackage{float}
\usepackage{caption}

\usepackage[utf8]{inputenc}

\preprint{FERMILAB-PUB-21-012-T}

\title{Systematically Testing Singlet Models for $\g$}

\author[a,b]{Rodolfo Capdevilla,}
\author[a]{David Curtin,}
\author[c,d]{Yonatan Kahn,}
\author[e,f,g]{Gordan Krnjaic}

\affiliation[a]{Department of Physics, University of Toronto, Canada}
\affiliation[b]{Perimeter Institute for Theoretical Physics, Waterloo, Ontario, Canada} 
\affiliation[c]{University of Illinois at Urbana-Champaign, Urbana, IL,  USA}
\affiliation[d]{Illinois Center for Advanced Studies of the Universe, University of Illinois at Urbana-Champaign, Urbana, IL, USA}
\affiliation[e]{Fermi National Accelerator Laboratory, Batavia, IL, USA}
\affiliation[f]{University of Chicago, Department of Astronomy and Astrophysics, Chicago, IL, USA}
\affiliation[g]{Kavli Institute for Cosmological Physics, University of Chicago, Chicago, IL, USA}

\emailAdd{rcapdevilla@perimeterinstitute.ca}
\emailAdd{dcurtin@physics.utoronto.ca}
\emailAdd{yfkahn@illinois.edu}
\emailAdd{krnjaicg@fnal.gov}

\date{\today}

\abstract{
We comprehensively study all viable new-physics scenarios that resolve the muon $(g-2)_\mu$ anomaly with only Standard Model singlet particles coupled to muons via renormalizable interactions. Since such models are only viable in the MeV -- TeV mass range and require sizable muon couplings, they predict abundant accelerator production through the same interaction that resolves the anomaly. We find that a combination of fixed-target (NA64$\mu$, $M^3$), $B$-factory (BABAR, Belle II), and collider (LHC, muon collider) searches can cover nearly all viable singlets scenarios, independently of their decay modes. In particular, future muon collider searches offer the only certain test of singlets above the GeV scale, covering all higher masses up to the TeV-scale unitarity limit for these models.  Intriguingly, we find that $\mathcal{O}(100~\mathrm{GeV})$ muon colliders may yield better coverage for GeV-scale singlets compared to TeV-scale concepts, which has important implications for the starting center-of-mass energy of a staged muon collider program.
}
\begin{document}

\maketitle

\section{Introduction}
\label{s.intro}

The Fermilab Muon $g-2$ collaboration has recently released its first measurement
of the muon's anomalous magnetic moment~\cite{Muong-2:2021vma,Muong-2:2021ojo,Muong-2:2021ovs,Muong-2:2021xzz}. Their new results  are consistent with previous Brookhaven measurements~\cite{Muong-2:2006rrc}, so the world average for $a_\mu \equiv \frac{1}{2} (g-2)_\mu$ now deviates from its Standard Model (SM) phenomenological prediction \cite{Aoyama:2020ynm,Aoyama:2012wk,Aoyama:2019ryr,Czarnecki:2002nt,Gnendiger:2013pva,Davier:2017zfy,Keshavarzi:2018mgv,Colangelo:2018mtw,Hoferichter:2019mqg,Davier:2019can,Keshavarzi:2019abf,Kurz:2014wya,Melnikov:2003xd,Masjuan:2017tvw,Colangelo:2017fiz,Hoferichter:2018kwz,Bijnens:2019ghy,Colangelo:2019uex,Pauk:2014rta,Danilkin:2016hnh,Jegerlehner:2017gek,Knecht:2018sci,Eichmann:2019bqf,Roig:2019reh,Colangelo:2014qya} by
\be
\label{amu-exp}
\Delta a_\mu = (251 \pm 59) \times 10^{-11},
\ee
which yields a statistically significant $4.2 \,\sigma$ discrepancy and may be the first laboratory evidence of physics beyond the 
SM (BSM).\footnote{Lattice calculations of the Hadronic light-by-light contribution to $a_\mu$ \cite{Gerardin:2019vio,Blum:2019ugy,Chao:2021tvp} as well as previous  lattice calculations of the Hadronic Vacuum Polarization (HVP) \cite{FermilabLattice:2017wgj,Budapest-Marseille-Wuppertal:2017okr,RBC:2018dos,Giusti:2019xct,Shintani:2019wai,FermilabLattice:2019ugu,Gerardin:2019rua,Aubin:2019usy,Giusti:2019hkz} are in agreement with the phenomenological values. However, a more recent calculation of the HVP from the BMW collaboration \cite{Borsanyi:2020mff} implies a value of $a_\mu$ that is consistent with the measured value. This calculation might be in tension with the SM electroweak fit \cite{Passera:2008jk,Crivellin:2020zul,Keshavarzi:2020bfy}, for which future lattice calculations and improved $R$-ratio data will likely be needed to clarify this situation.}  If the anomaly is due to BSM states, their presence induces the effective operator
 \be
 \label{leff}
 {\cal L}_{\rm eff} = 
 C_\mathrm{eff} \frac{v}{\Lambda^2}  ({\mu}_L  \sigma^{\nu \rho}  \mu^c )  F_{\nu \rho}   + {\rm h.c.}~,
 \ee
 where $\mu_L$ and $\mu^c$ are the muon Weyl spinors, $v = 246$ GeV is the Higgs vacuum expectation
 value (VEV), $C_\mathrm{eff}$ is a dimensionless coefficient and $\Lambda$ is a BSM mass scale. This effective operator generates a contribution $a_\mu^{\rm BSM}$ to $\g$ that can in principle resolve the anomaly by setting $a_\mu^{\rm BSM} = \Delta a_\mu$.
 If the BSM states are charged under the electroweak (EW) gauge group, there is a vast landscape of models that can resolve the anomaly (see Refs. \cite{Capdevilla:2020qel,Capdevilla:2021rwo,Athron:2021iuf} for a review), hereafter referred to as {\it electroweak (EW) models}. Such models tend to
  have multiple free parameters and can, in principle, span a mass range for $\Lambda$ anywhere between $\sim$ 100 GeV$-$100 TeV~\cite{Capdevilla:2020qel,Capdevilla:2021rwo,Athron:2021iuf,Dermisek:2013gta,Freitas:2014pua,Queiroz:2014zfa,Calibbi:2018rzv,Dermisek:2021ajd,Arkani-Hamed:2021xlp,Dermisek:2021mhi}.

However, if the BSM states that generate \Eq{leff} are singlets under the SM gauge group ({\it singlet models}), the Higgs VEV and chirality flip in \Eq{leff} must both arise from the muon mass, so \Eq{leff} becomes proportional to $m_\mu$,
 \be
 \label{leff2}
 {\cal L}_{\rm eff} =  C^\prime_{\rm eff}   \frac{m_\mu}{M^2}  ({\mu}_L  \sigma^{\nu \rho}  \mu^c )  F_{\nu \rho}   + {\rm h.c.}~,
 \ee
 where $M$ is the mass of a SM singlet particle that has been integrated out to generate these operators.
  This feature greatly 
 simplifies the model landscape down to two renormalizable 
 interactions:
 \be
 \label{singlet-couplings}
 g_S S (\mu_L \mu^c~ + \mu^{c  \dagger} \mu_L^\dagger),~~~~
  g_V  V_\nu (\mu^\dagger_L \bar \sigma^\nu \mu_L + \mu^{c  \dagger}\bar \sigma^\nu \mu^{c})~,
  \label{lag-singlets}
 \ee
corresponding to BSM scalars ($S$) and vectors ($V$), respectively, and greatly compresses the parameter space compared to EW models because of the muon mass suppression.\footnote{Alternative possibilities involving 
 pseudo-scalars or axial vectors contribute the wrong sign to $a_\mu^{\rm BSM}$ and do not resolve the anomaly, and interactions whose leading contribution to $\g$ involves two-loop diagrams are severely constrained -- see Appendix \ref{appendixA} for a discussion.}
 
  \begin{figure}[t]
\center
\includegraphics[width=5in]{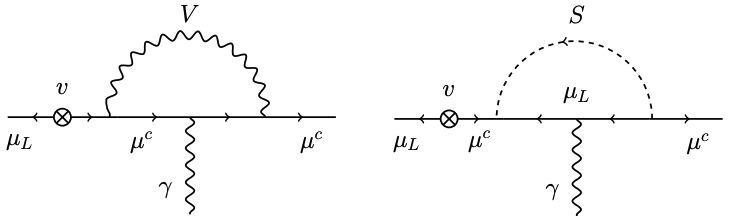}
\caption{One-loop contributions to $\g$ from vector (left) and scalar (right) singlets.
\label{f.feynmang-2}}
\end{figure}
  
 In this paper, we systematically analyze all singlet models that resolve the $\g$ anomaly and survey how numerous proposed experiments can cover much of the remaining viable parameter space. Every singlet model can be uniquely specified by a dimensionless coupling $g_{S,V}$, a corresponding singlet mass $m_{S,V}$, and a possible additional decay width of the singlet to other non-muonic particles. Thus, singlet models are generically more predictive than electroweak models and can be studied without specifying many model-dependent features (for instance, the particle spectra of supersymmetric models). However, because the scalar singlet interaction in Eq.~(\ref{lag-singlets}) is not gauge-invariant under the SM, it \emph{must} be UV-completed, and the extra particles required for this UV completion will play an important role in further restricting the parameter space of the scalar singlet model once constraints from e.g.\ electroweak precision observables are taken into account.

This paper is organized as follows. In Section \ref{s.singlets} we summarize
  the landscape of  singlet models and identify the allowed parameter space in each scenario, which is bounded from below by cosmology and from above by unitarity. In Sections \ref{s.ft}, \ref{s.bfac}, and \ref{s.colliders} we discuss future prospects for
  fixed target experiments, $B$-factories, and collider experiments, respectively. We conclude in Section \ref{s.conclusions} with an outlook on  the types of experiments needed to fill any remaining gaps in parameter space to conclusively test singlet explanations for $(g-2)_\mu$. Our main results are summarized in \fref{babar_singlet}. The Appendices contain discussions of 2-loop models, searches for singlets with non-muonic visible decays, and UV completions of the scalar singlet models.

\section{Singlet Models}
\label{s.singlets}

 Throughout this paper, {\it Singlet Models} refers to the family of models where  one of the interactions in \Eq{lag-singlets} generates $\Delta a_\mu$ in \eref{amu-exp}.
 In this section, we outline the general properties of these models and identify their viable parameter space for resolving the 
 $\g$ anomaly.

 \subsection{Vector Singlets}
 \label{s.singlets.vector}
 For a vector singlet with no axial couplings, the partial decay width to di-muons is 
 \be
 \label{GammaV}
 \Gamma(V \to \mu^+\mu^-) = \frac{g_V^2 m_V}{12\pi}\left(  1 + \frac{2m_\mu^2}{m_V^2}     \right)\sqrt{1- \frac{4m_\mu^2}{m_V^2}}~,
 \ee
for $m_V > 2m_\mu$. The vector contribution to $\g$ from Fig. \ref{f.feynmang-2} (left) can be
 written
\be
\label{aV}
a^{V}_{\mu}  =    \frac{g_V^2 }{4\pi^2} \int_0^1\! dz  \frac{   z (1-z)^2}{  (1-z)^2 +    z (m_V/m_\mu)^2} ~,
\ee
and the favored parameter space for resolving the $\g$ anomaly is shown in the orange band of Fig.~\ref{f.g-2contours}.
In principle, there are many possible choices for abelian SM extensions that contain a vector-muon interaction. However, for anomaly-free U(1) gauge extensions to the SM, the favored parameter
space for resolving $\g$ is already excluded in most variants. 

\begin{itemize}
\item{\bf Kinetically-mixed dark photon} \\
The popular kinetically-mixed ``dark photon'' model features a new vector singlet corresponding to a secluded
U(1) gauge group under which SM fermions carry no charges \cite{Pospelov:2008zw}. In this scenario the muon-dark photon coupling is induced through kinetic mixing between dark and visible photons, so the dark photon couples to all charged fermions with equal strength.
Consequently, this model is excluded as a solution to $\g$ by a variety of $B$-factory, beam dump, and  rare meson decay searches -- see \cite{Ilten:2018crw,Bauer:2018onh} for a summary of constraints.  This conclusion holds regardless of whether the dark photon decays visibly or invisibly.\footnote{ However, there is a limited parameter space for exotic decays that yield both visible and invisible particles via dark sector cascades in each decay event and may be tested with dedicated $B$ factory and fixed target searches \cite{Mohlabeng:2019vrz}.}

\begin{figure}[t]
\center
\includegraphics[width=2.8in]{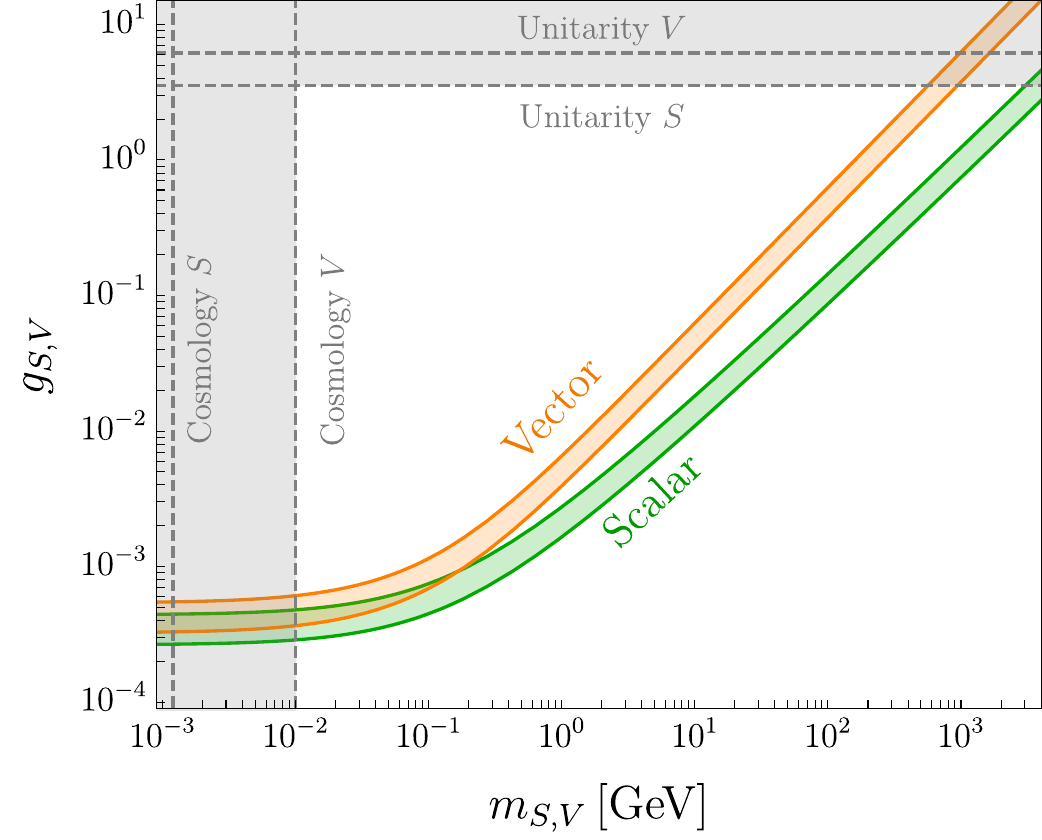}~
\includegraphics[width=2.8in]{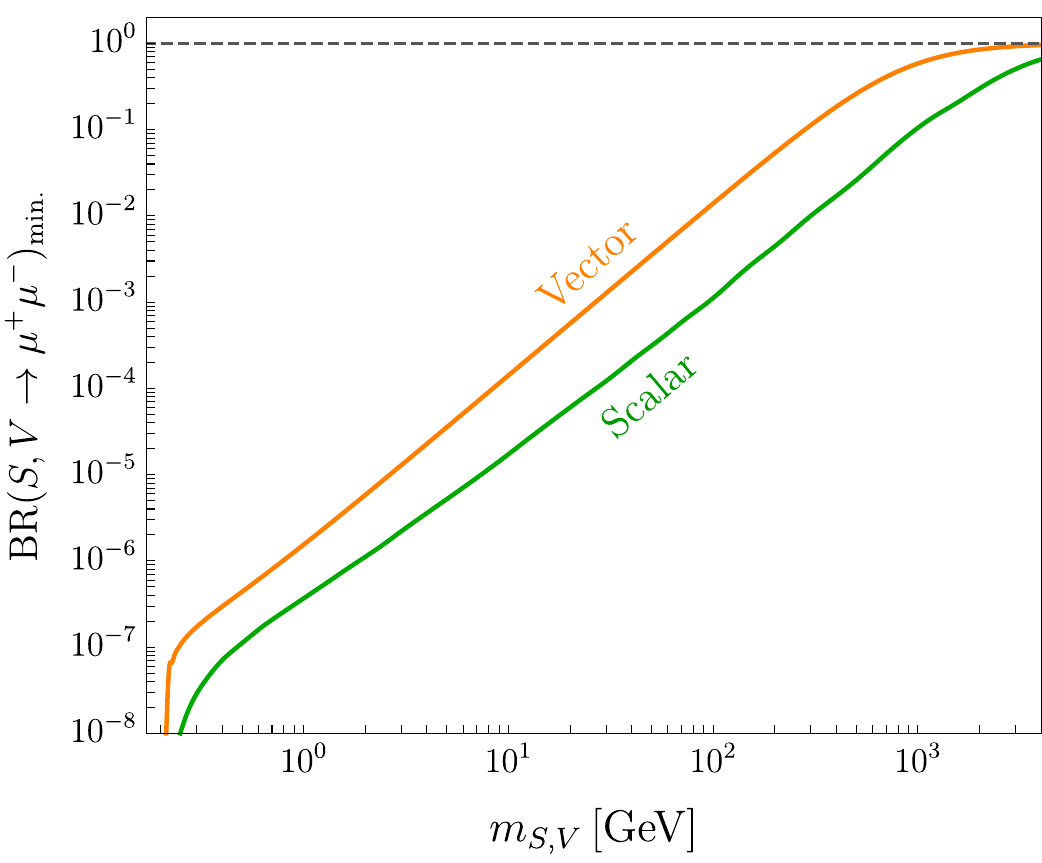} 
\caption{{\bf Left:} Parameter space for which one SM singlet scalar or  vector particle resolves the $\g$ anomaly. The thickness
of each band represents the $\pm 2\sigma$ confidence interval and the vertical axis is the corresponding muon-singlet coupling from \Eq{singlet-couplings}. The vertical shaded region represents the bound on light relativistic species present in equilibrium during big bang nucleosynthesis. The horizontal shaded region is the bound from partial wave unitarity. For a discussion 
of these bounds, see Section \ref{massrange}.
{\bf Right:} Minimum di-muon branching fraction for vector and scalar couplings that resolve the 
$\g$ anomaly, from Eqs.~(\ref{branch-min-vector}) and~(\ref{branch-min-scalar}). 
\label{f.g-2contours}
}
\end{figure}


\item{\bf Anomaly-free U(1) extensions} \\
Minimal abelian extensions to the SM promote an anomaly-free linear combination of baryon $(B)$, lepton $(L)$, and lepton flavor $(L_i)$ quantum numbers, where $i = e, \mu,\tau$, to a gauge symmetry.
Some examples include $B-L$, $B - 3 L_i$, $L_i - L_j$. Since nearly all of these options include some coupling to first-generation SM fermions, they are subject to the same constraints that apply to the dark photon and are, therefore, also excluded \cite{He:1990pn,Dror:2017nsg,Ballett:2019xoj} (unless new fields are introduced \cite{Allanach:2015gkd,Raby:2017igl,Kawamura:2019rth,Bodas:2021fsy,Mondal:2021vou,Greljo:2021xmg,Lee:2021gnw}).
A well known exception is gauged $L_\mu- L_\tau$ \cite{Altmannshofer:2014pba,Escudero:2019gzq,Krnjaic:2019rsv,Holst:2021lzm}, which has no tree-level couplings to first-generation particles and evades many of the bounds that arise from couplings to protons and electrons. This model is viable for vector masses in the $\sim$ 15 -- 210 MeV range \cite{Escudero:2019gzq} in which $V$ decays exclusively to neutrinos and can be tested with low-energy fixed-target experiments -- see Sec.~\ref{s.ft} for more details.

In the case of the most general $U(1)_X$ extension for $\g$ including right-handed neutrinos (needed for anomaly cancellation) one can have a vast family of models \cite{Greljo:2021npi}. These can be classified as dark photon-like or  $L_\mu- L_\tau$-like in case they include (former) or not (latter) interactions with first generation leptons. Most of these models are either excluded or highly constrained \cite{Greljo:2021npi}.

\item{\bf Anomalous U(1) extensions}\\
Since most bounds on anomaly-free U(1) models involve constraints on their non-muonic interactions, in principle it 
is possible to open up parameter space by coupling the vector only to muons. However, such an approach introduces interactions that grow with energy via non-decoupling triangle diagrams that amplify otherwise rare SM processes \cite{Dror:2017nsg,Kahn:2016vjr,Dror:2017ehi}. While such models have not been studied exhaustively in the context
of explaining $\g$, it is expected that they would be strongly constrained by existing bounds (e.g. 
rare meson decays involving muon final states). We note that a fully unitary theory with anomalous interactions
at low energies must ultimately invoke additional SM charged states to cancel triangle diagrams
at some scale; however, there is considerable model dependence in how this is achieved.

\end{itemize}

For the remainder of this paper, we will take an agnostic approach towards the models that realize the
vector-muon coupling, while noting that specific realizations of vector singlets might have additional bounds beyond those that 
we present. Taking into account only the irreducible coupling to muons which generates $\g$, we will present a phenomenological treatment that studies only di-muon decays or invisible decays,
\begin{align}
m_V < 2m_\mu: & \ V\to \displaystyle{\not}{E} & {\rm (BR = 1)}~,\\
m_V > 2m_\mu: & \ V\to \mu^+ \mu^-  & {\rm (BR = 1 \ or \ BR = min.)}~,
\end{align}
where $\displaystyle{\not}{E}$ also allows for a long-lived $V$ that effectively gives rise to missing energy signatures. For $m_V > 2m_\mu$ in this agnostic approach, the \emph{total} width is bounded by unitarity at $\Gamma_{V,\rm max} \lesssim \frac{1}{2} m_V$ \cite{Schwartz:2014sze}.  Saturating this maximum width, even if the dominant $V$ branching fraction is into other channels (visible or invisible), the {\it minimum} vector branching fraction to di-muons is 
\be
\label{branch-min-vector}
{\rm BR}(V \to \mu^+ \mu^-)_{\rm min}  \approx \frac{g_V^2}{6\pi} 
\approx 5 \times 10^{-2} \brac{m_V}{200 \, \rm GeV}^2~,
\ee
where we have used \Eq{GammaV} in the $m_V \gg m_\mu$ limit and
expressed \Eq{aV} in terms of the central value in \Eq{amu-exp}; this minimum value applies
to any vector singlet that resolves the $\g$ anomaly.
In Fig. \ref{f.g-2contours} (right panel) we show the minimum branching fraction as a function of singlet mass. 

As an example of how this general analysis becomes constrained in the context of particular models, if $V$ is the gauge boson of an anomaly-free group involving $L_\mu$, then the only viable parameter space is for $m_V < 2m_\mu$ with predominantly invisible $V \to \bar \nu\nu$ decays, since $m_V > 2m_\mu$ is excluded by the irreducible visible decay channel. Other channels due to, for example, a small kinetic mixing between $V$ and the photon, are strongly suppressed.

  \subsection{Scalar Singlets }\label{s.scalar_singlets}
  For scalar singlets, the partial decay width to di-muons is
  \be
  \label{scalar-width}
  \Gamma(S \to \mu^+ \mu^-) = \frac{g_S^2 m_S}{8\pi} \left(  1 - \frac{4m_\mu^2}{m_S^2}  \right)^{3/2}~,
  \ee
for $m_S > 2m_\mu$, and the corresponding contribution to $\g$ from Fig. \ref{f.feynmang-2} (right) can be written
\be
\label{aS}
a^{S}_{\mu}  =    \frac{g_S^2}{8\pi^2} \int_0^1 dz \frac{ (1+z) (1-z)^2}{(1-z)^2 +  z (m_S/m_\mu)^2 }~~.
\ee
The favored parameter space for resolving the $\g$ anomaly is shown in the green band of Fig. \ref{f.g-2contours}.
 This contribution is comparable to the corresponding expression for vector singlets in \Eq{aV}. 
 
 Unlike in the vector case, the scalar Yukawa interaction $S \mu_L \mu^c$  in \Eq{lag-singlets}   is not gauge-invariant under SU(2)$_L \times$ U(1)$_Y$ and must 
 arise from a higher-dimension operator. 
 In principle, this could be avoided if $S$ mass-mixes with the SM Higgs. However, the SM Higgs-muon Yukawa coupling is $y_\mu \approx 6 \times 10^{-4}$, so from Fig. \ref{f.g-2contours}, such a contribution is too small for all but the lightest $S$ masses, even before taking into account strong suppression from the singlet mass or its tiny $(\ll 1)$ mixing angle with the Higgs \cite{Krnjaic:2015mbs}. 
Thus, the lowest-dimension operator that can generate a flavor-specific coupling to muons is 
 \be
 \label{singlet-yukawa-higher-dim}
 {\cal L} \supset \frac{1}{\Lambda} S H^\dagger L \mu^c + {\rm h.c.} \to  \frac{v}{\Lambda} S \mu_L \mu^c + {\rm h.c.} ~,
 \ee
 where $L$ is the second-generation lepton doublet and $\Lambda$ is a BSM mass scale generated by heavier particles that are integrated out to generate the singlet-muon Yukawa coupling after electroweak symmetry breaking.  
 
We defer a detailed discussion of UV completions for the scalar singlet scenarios to \sref{EWPT} and \aref{appendixB}. Here we only point out that since scalar Yukawa interactions are not constrained by anomaly cancellation requirements, the flavor structure of these interactions is only constrained by experimental considerations:

 \begin{itemize}
 \item{\bf Mass Proportional Couplings:}\\
 As noted above, $S$ cannot mass-mix with the SM Higgs because the 
 muon Yukawa is too small to explain $\g$, but a larger muon coupling can be achieved if $S$ 
 is part of a two-Higgs doublet model with leptophilic couplings. In such a model, the scalar couplings
 are proportional to charged lepton masses and can be larger than their nominal SM values (see \cite{Batell:2016ove,Chen:2017awl} for an example). However, the parameter space for resolving the $\g$ anomaly within this model has been excluded by the BABAR experiment in a null search for BSM $e^+e^- \to \tau^+\tau^- \mu^+\mu^-$ events \cite{BaBar:2020jma}.

\item{\bf Flavor Specific Couplings:} \\
The interaction in \Eq{singlet-yukawa-higher-dim} can also arise by coupling the singlet to 
new electroweak-charged fermions that mix with the muon, followed by integrating them out at some scale $\Lambda$. 
In principle, such an approach allows any flavor structure (subject to empirical constraints) and many options have been considered in the literature \cite{Chen:2015vqy,Batell:2017kty,Hiller:2019mou,Egana-Ugrinovic:2019wzj,Liu:2020qgx,Hiller:2020fbu,Bissmann:2020lge,Batell:2021xsi}.
\end{itemize}

Thus, for the remainder of this paper, we will not assume any particular flavor structure in the $S$ couplings. For light scalars below the di-muon threshold, we will consider 
\be
m_S < 2m_\mu: \ S \to ~ \gamma\gamma, \ \displaystyle{\not}{E} \qquad {\rm (BR = 1)}~.
\ee 
As before, $\displaystyle{\not}{E}$ represents invisible decays or a long-lived $S$. Note that for singlet scalars, the di-photon coupling arises at one-loop through the $S$ coupling to muons, which is required for $\g$.\footnote{Of course, this contribution can be tuned against an additional higher-dimension term $\frac{1}{\Lambda'} F_{\mu \nu}F^{\mu \nu} S$ present in the Lagrangian, but this term would require its own UV completion, and fine-tuning the photon coupling away entirely requires an exquisite coincidence of scales.} As we will see in Sec.~\ref{beam-dump}, the decay length assuming only the $\gamma \gamma$ channel is macroscopic, such that it may manifest as missing energy $\displaystyle{\not}{E}$ in a collider experiment but be visible at a beam-dump experiment. Note also that the $\gamma \gamma$ channel is forbidden for vectors by the Landau-Yang theorem. Since $S \to e^+ e^-$ is not required by $\g$ and depends on the flavor structure of the model, in our model-agnostic treatment we will not consider it in detail in this work, but we briefly comment on this channel in \aref{Bfactoryappendix}. For scalars above the di-muon threshold, 
we will consider the irreducible decay channel
\be
m_V > 2m_\mu: & \ S\to \mu^+ \mu^-  \qquad {\rm (BR = 1 \ or \ BR = min.)}~.
\ee
As in the vector case discussed above, the minimum muon couplings required for $\g$ and the unitarity upper limit on the singlet's total width $\Gamma_{S,\rm max} \approx m_S/2$ 
imply a {\it minimum} di-muon branching fraction independently of other decay channels
\be
\label{branch-min-scalar}
{\rm BR}(S \to \mu^+ \mu^-)_{\rm min}  \approx \frac{g_S^2}{4\pi}
\approx 4\times 10^{-3} \brac{200 ~\rm GeV}{m_S}^2~,
\ee
where we have used \Eq{scalar-width} in the $m_S \gg m_\mu$ limit and 
followed the same logic that leads to \Eq{branch-min-vector} in the vector case.
In Fig. \ref{f.g-2contours} (right panel), we show the minimum branching fraction as a function of singlet mass.

 \begin{figure}[t]
\center
\vspace*{-10mm}
\hspace{-0.75cm}
\includegraphics[width=2.9in]{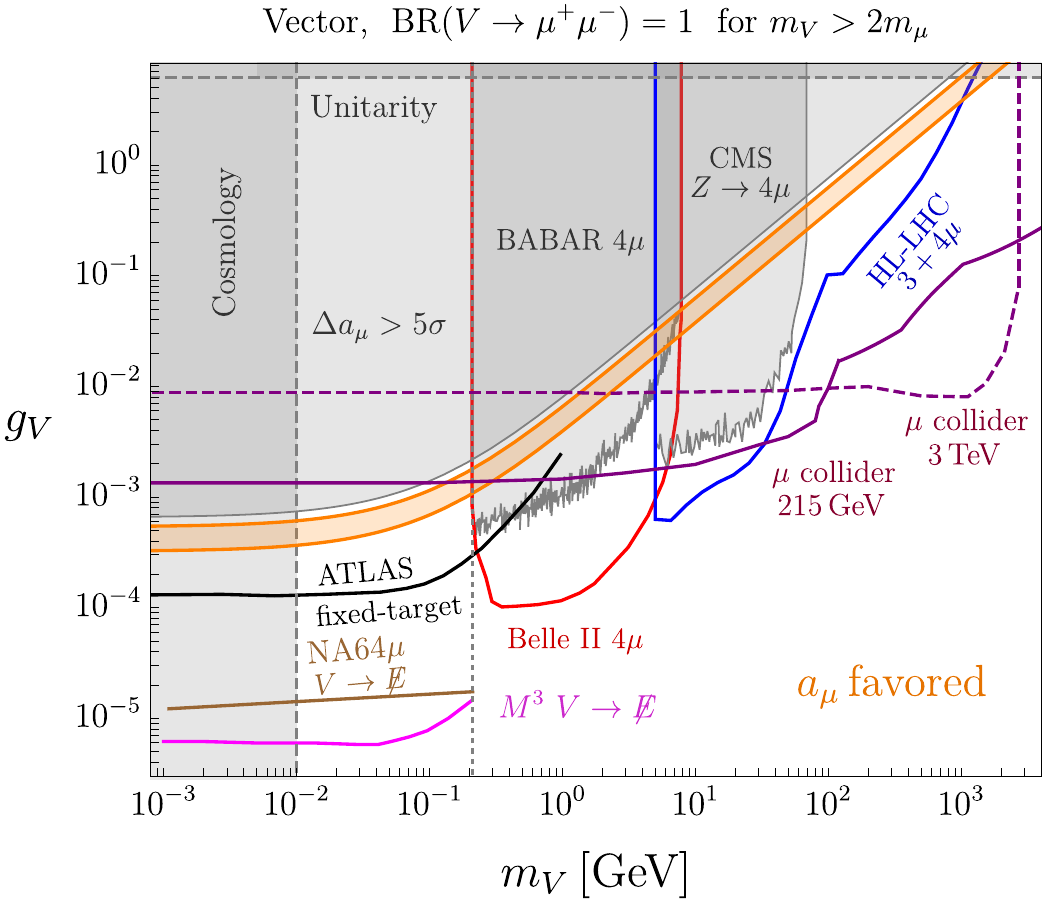} ~
\includegraphics[width=2.9in]{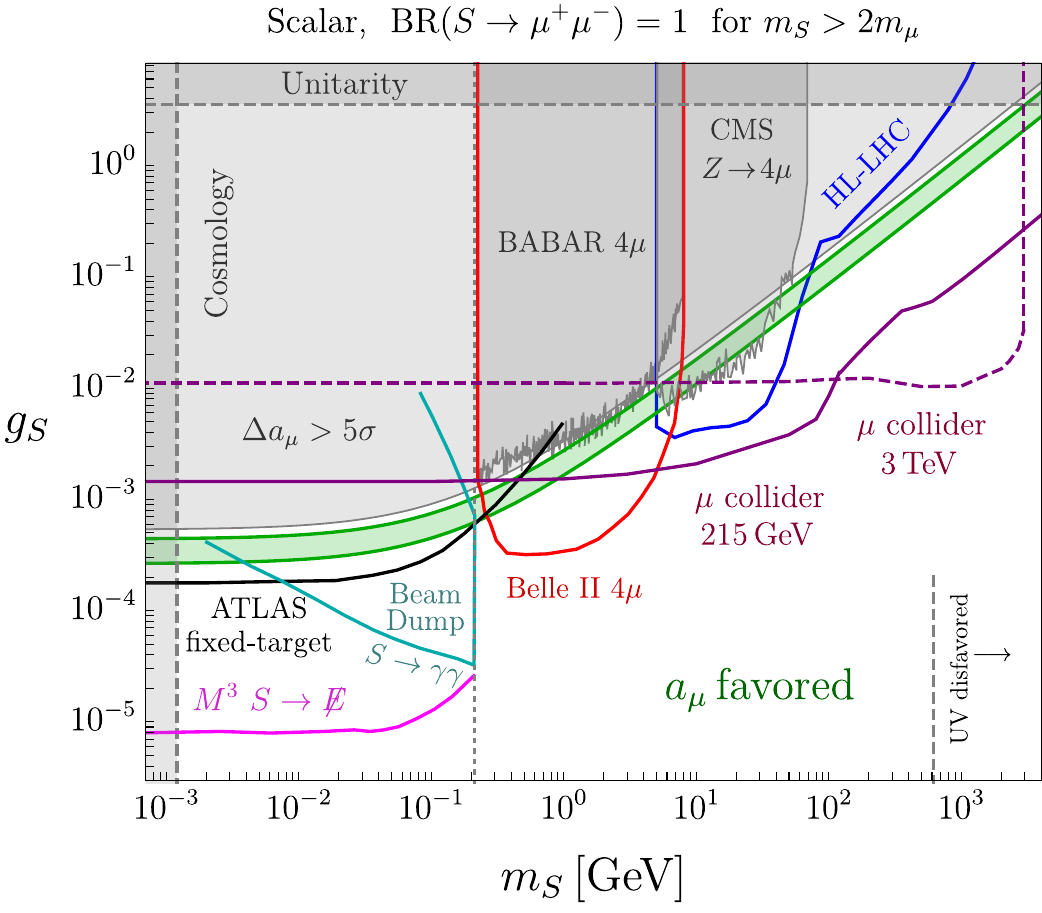}  \\ 
\hspace{-0.75cm}
\includegraphics[width=2.9in]{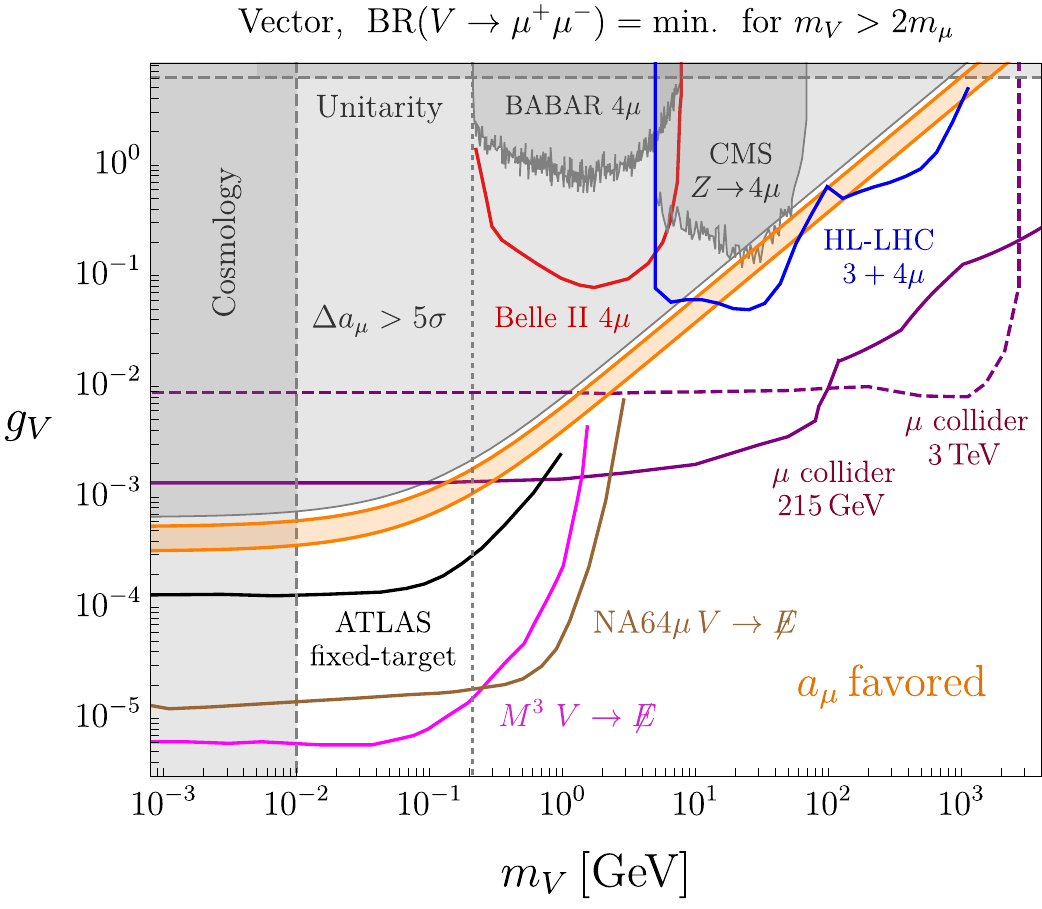} ~
\includegraphics[width=2.9in]{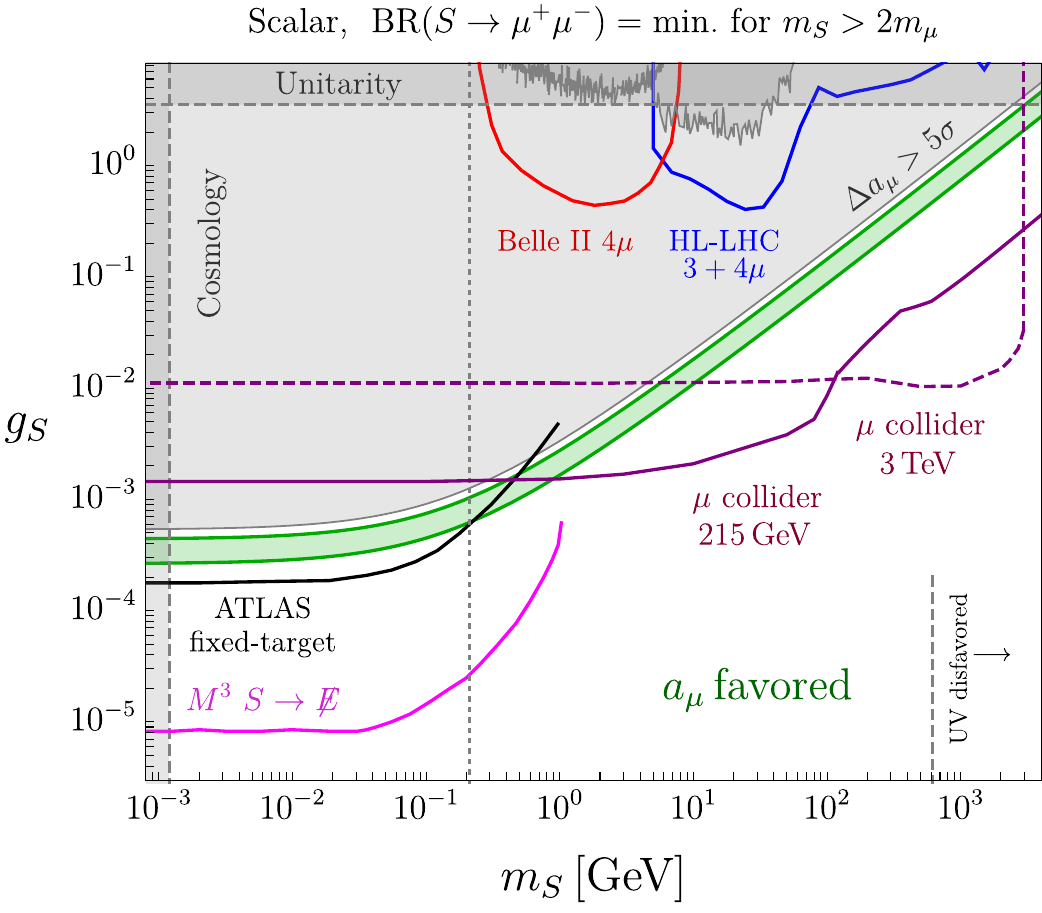} 
\caption{
{\bf Top:}
Limits and projections on muon-philic vector (left) and scalar (right) singlets,
assuming only di-muon decays where kinematically allowed. The green/orange bands represent the parameter space that resolves $\g$. Existing experimental limits are shaded in gray (Supernova constraints, not shown, can probe scalar masses up to 20 MeV and couplings up to $4\times10^{-3}$ \cite{Caputo:2021rux}). Projections are indicated with colored lines. The  $M^3$ \cite{Kahn:2018cqs}, NA64$\mu$ \cite{Gninenko:2018ter}, and ATLAS fixed-target \cite{Galon:2019owl} experiments 
probe invisibly-decaying singlets; projections here assume a 100\% invisible branching 
fraction (see Sec.~\ref{s.ft}). The BABAR limits and Belle II projections are computed 
following the procedure described in Sec.~\ref{s.bfac}. The LHC limits and HL-LHC projections in the $3\mu$/$4\mu$ channels, along with the mass range disfavored by UV completions for scalar singlets, are discussed in Sec.~\ref{s.colliders} and \aref{appendixB}.
The purple muon collider projections are based on proposed analyses~\cite{Capdevilla:2021rwo} reviewed 
in Sec. \ref{MuC}.
For scalar singlets whose width is determined entirely by the muon coupling (top right), we also show projections for a $S \to \gamma\gamma$ beam
dump search \cite{Chen:2017awl} on the minimal assumption that the scalar-photon 
coupling arises from integrating out the muon as discussed in Sec. \ref{beam-dump}.
{\bf Bottom:} Same as the top row, only here we assume that for $m_{S,V}  > 2m_\mu$, the singlets
have the {\it minimum} di-muon branching fraction consistent with unitarity 
using {\rm Eqs.}~(\ref{branch-min-vector}) and (\ref{branch-min-scalar}). The curves which
are unaffected by this change of muonic branching fraction correspond to searches that are insensitive to the singlet's decay modes. Projections for $M^3$, NA64$\mu$, and ATLAS fixed-target experiments assume a $\simeq 100\%$ invisible branching fraction for $m_{S/V} > 2m_\mu$, which is model-dependent.
 \label{babar_singlet}}
\end{figure}

\subsection{Viable Mass Range}
\label{massrange}
Although the couplings required to explain $\g$ for light singlets are feeble from 
a collider perspective, they nonetheless suffice to thermalize these new states with the SM radiation bath in the early universe, provided
the initial temperature exceeds a given choice of singlet mass. For light singlets ($m_{S,V}\ll T$), the early universe production and annihilation rates scales as $\Gamma_{S,V} \sim g_{S,V}^2 T$, where $T$ is the temperature of the radiation bath. Comparing this rate to Hubble expansion implies
thermalization for temperatures 
\be
T \gtrsim  \frac{ 1.66 \sqrt{g_\star} }{g_{S,V}^2 M_{\rm Pl} } \approx 1\, \text{eV} \brac{5 \times 10^{-4}}{g_{S,V} }^2~,
\ee
where $g_\star$ is the number of relativistic species in the bath and $M_{\rm Pl} = 1.22 \times 10^{19}$ GeV is the Planck mass. 
Thus, for singlet masses below the MeV scale with couplings that resolve the $\g$ anomaly 
in Fig. \ref{f.g-2contours} (left panel), there will be a thermal $\propto T^3$ number density of singlets in the early universe at temperatures $T \sim \ {\rm MeV}$, which
 increases the expansion rate during Big Bang Nucleosynthesis (BBN) and spoils the successful SM prediction of 
light element yields.  Avoiding the BBN bound requires \cite{Escudero:2019gzq,Krnjaic:2019dzc}
\be
m_V \gtrsim 10 \, \text{MeV}~~,~~ m_S \gtrsim 1 \, \text{MeV}~,
\ee
which are represented in the shaded regions labeled ``cosmology'' in Figs. \ref{f.g-2contours} and \ref{babar_singlet}.

For singlets above the muon mass, the BSM contribution to $\g$ scales as $a_\mu^{\rm BSM} \propto g_{S,V}^2/m_{S,V}$ so, in principle, any choice of mass can resolve the anomaly. However, for sufficiently large couplings,
singlet interactions violate partial wave unitarity. As shown in~\cite{Capdevilla:2021rwo}, imposing tree-level unitarity constraints on muonic Bhabha scattering $\mu^+ \mu^- \to \mu^+ \mu^-$ through an $s$- or $t$-channel $S/V$ yields  $g_S < \sqrt{4\pi}$ and $g_V < \sqrt{12\pi}$. This implies an upper bound of $m_S < 2.7 \ {\rm TeV}$ and $m_V < 1.1 \ {\rm TeV}$, or a UV completion involving additional states at the same mass scale as the singlet itself. The combination of cosmology and unitarity constraints provides a robust two-sided boundary to the viable mass range for singlet solutions to $(g-2)_\mu$:
\be
\text{few} \ {\rm MeV} \lesssim m_{S,V} \lesssim  \text{few} \ {\rm TeV}.
\ee
In Fig.~\ref{babar_singlet}, we show the viable parameter space for both visibly- and invisibly-decaying $S$ and $V$, along with constraints and projections from fixed-target experiments, $B$-factories, and colliders, which we discuss in detail in the following sections.

\section{Fixed Targets}
\label{s.ft}

\begin{figure}[t]
\center
\includegraphics[width=5in]{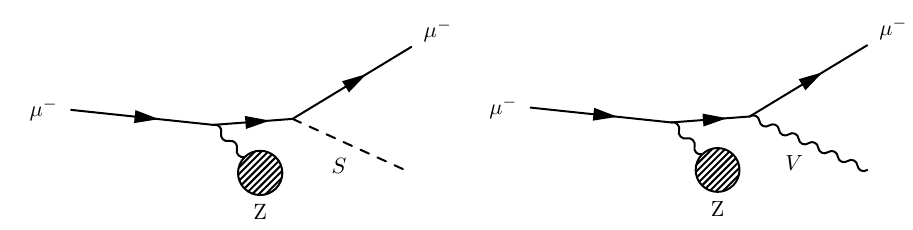}
\caption{Radiative singlet production via coherent muon-nucleus scattering at 
muon beam fixed target experiments $M^3$ \cite{Kahn:2018cqs}, NA64$\mu$ \cite{Gninenko:2018ter}, and proposed beam dump searches \cite{Chen:2017awl}.
\label{f.fixed-target}}
\end{figure}

Fixed-target experiments are a promising strategy to search for new light ($<$ GeV) weakly-coupled SM singlet particles in 
a variety of contexts
\cite{Altmannshofer:2014pba,Chen:2017awl,Kahn:2018cqs,Battaglieri:2017aum,Izaguirre:2013uxa,Berlin:2018bsc,BDX:2019afh,Berlin:2018pwi,NA64:2017vtt,Proceedings:2012ulb,Blumlein:2011mv,Celentano:2014wya,Gninenko:2014pea,Gninenko:2019qiv,Kahn:2012br,Balewski:2014pxa,Batell:2014mga,Izaguirre:2014bca}, including the idea of using the ATLAS calorimeters as a fixed-target \cite{Galon:2019owl}.
As discussed in Sec.~\ref{s.singlets}, viable models for $\g$ with light singlets  require
muon-philic interactions, but most fixed-target experiments involve beams of electrons or protons and
are, therefore, insensitive to the remaining parameter space that resolves the anomaly. However, several
new muon-beam fixed-target experiments have recently been proposed to address this shortcoming 
 by exploiting singlet production in muon-nucleus scattering, see Fig. \ref{f.fixed-target}.

\subsection{Beam Dump Searches for $S \to \gamma\gamma$}
\label{beam-dump}

In a beam dump experiment, a high-intensity muon beam impinges on
a large, dense target which is surrounded by shielding material to block SM particles. 
If a long-lived, visibly-decaying singlet is produced in the dump, it can travel 
unimpeded through the shielding and decay to visible SM particles 
in a downstream detector.

Such a setup can cover nearly all of the remaining parameter
space for scalar singlets with $m_S < 2m_\mu$ that decay via $S \to \gamma \gamma$, induced at one-loop through 
the minimal scalar-muon Yukawa coupling required to address the  $\g$ anomaly \cite{Chen:2017awl}:
\be
\hspace{1cm} g_S S \mu_L \mu^c ~~\longrightarrow~~   c \frac{\alpha }{16 \pi^2} \frac{ g_S}{ m_\mu} S F^{\mu\nu}F_{\mu\nu}~ ~~~(E \ll m_\mu),
\ee
where $c$ is an order-one number and in the second expression the muon has been integrated out. If the $S$-$\mu$ Yukawa coupling is the only contribution 
to this operator, then for $g_S$ that resolve $\g$ anomaly, the decay length is \cite{Chen:2017awl}
\be
\label{decay-length}
L_S = 20 \, {\rm m}\, \brac{E_S}{3\, \rm GeV} \brac{5 \times 10^{-4}}{g_S}^2 \brac{100 \, \rm MeV}{m_S}^4,
\ee
where $E_S$ is the energy of the $S$ particle and the fiducial values for $g_S$ and $m_S$ are chosen 
to match the favored $\g$ parameter space in Fig.~\ref{f.g-2contours} (left panel).

In the top-right plot of Fig.~\ref{babar_singlet} we show the projections from Ref.~\cite{Chen:2017awl} for a proposed beam dump search for long lived particles with a 3 GeV beam  and $3 \times 10^{14}$ muons on target, assuming only the minimal muon-induced coupling to di-photons (Model B from Ref.~\cite{Chen:2017awl}). However, note that if other model-dependent decay processes are introduced, the singlet  $S$ 
might not be a long-lived particle and this search strategy would lose sensitivity. Similar sensitivity could
be achieved for the $S \to e^+e^-$ channel if the singlet-electron coupling is chosen to reproduce the same 
decay length as in \Eq{decay-length}, but unlike the minimal muon-induced di-photon decay, this displacement scale
does not automatically follow from the same coupling that resolves the $\g$ anomaly (also see \aref{Bfactoryappendix} for further discussion). 

\subsection{Missing Momentum Searches for $S/V \to \displaystyle{\not}{E}$} 
\label{missing-energy/momentum}
The proposed NA64$\mu$ \cite{Gninenko:2014pea,Gninenko:2019qiv} and 
$M^3$ \cite{Kahn:2018cqs} experiments aim to test the remaining $\g$ parameter
space for invisibly-decaying singlet particles. In these experiments, 
a muon beam of moderate intensity impinges on an active target, which monitors the beam energy as it passes
through. Signal events are those where the outgoing muon energy is reduced by $\mathcal{O}(50\%)$ and no other SM particles are observed  in downstream veto detectors; the limiting factor for the luminosity is the requirement to only have a single muon per bunch to mitigate pileup. In \fref{babar_singlet}, we show the NA64$\mu$ and $M^3$ projections for  $10^{12}$ and  $10^{13}$ muons on target, respectively; these reference values match the ultimate reach projections from 
Refs. \cite{Gninenko:2014pea} and \cite{Kahn:2018cqs}, where Ref.~\cite{Gninenko:2014pea} only considered invisibly-decaying vectors. Note that these projections cover the remaining parameter space
that resolves the $\g$ discrepancy within the anomaly-free $L_\mu-L_\tau$ gauge extension discussed in 
Sec. \ref{s.singlets.vector}, where there is an order-1 invisible branching fraction to neutrinos for $m_V > 2m_\mu$. 
 Preliminary data is expected from NA64$\mu$ \cite{Sieber:2021fue} in the very near future.

\section{$B$-Factories}
\label{s.bfac}

\begin{figure}[t]
\center
\includegraphics[width=5in]{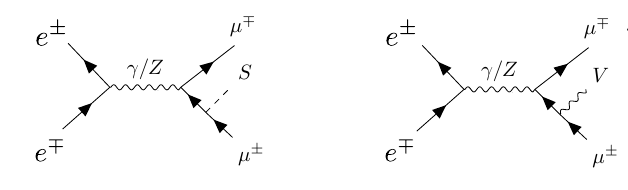} ~~~
\caption{Representative Feynman diagrams that yield singlet scalar (left) and vector (right) production at a $B$-factory via $e^+e^-$ annihilation. The BABAR 
and Belle II search strategies discussed in Sec. \ref{s.bfac} involve the $4\mu$ channel
in which $S/V \to \mu^+\mu^-$ decays yield $4\mu$ final states with an opposite-sign di-muon resonance that reconstructs the singlet's mass. 
}
\label{babar-feynman}
\end{figure}

We now discuss $B$-factory searches for singlets decaying to muons. Singlet models generate new contributions to $e^+e^- \to 4\mu$ events from $e^+ e^- \to \mu^+ \mu^- + S/V \to 4 \mu$ processes in which the singlet $S$ or $V$ is radiated 
from a final-state muon line and decays via $S/V \to \mu^+\mu^-$, as depicted in Fig. \ref{babar-feynman}.  This search strategy
can test the $\g$ favored parameter space for light singlets with appreciable
branching fractions to di-muons, but other search strategies are necessary for invisibly-decaying singlets. We briefly comment on other visible final-state searches at $B$-factories in \aref{Bfactoryappendix}.

The BABAR collaboration has performed a search for muon-philic vectors
with $L_\mu - L_\tau$ gauge interactions and excluded the 
$\g$ favored parameter space for masses between $2m_\mu$ and 10 GeV with 514 fb$^{-1}$ of data \cite{BaBar:2016sci}.
Since an $L_\mu-L_\tau$ gauge boson in this mass range can decay to muons, taus, and their corresponding
neutrino flavors, in Fig. \ref{babar_singlet} (left panel) we rescale their limit for our vector $V$ for which
 BR$(V \to \mu^+ \mu^-) = 1$; for models in which $V$ has a smaller branching fraction, the limit on $g_V$ gets weaker by 
 a factor of $\sqrt{  \text{BR}     (V \to \mu^+ \mu^-)}$. 

In the right panel of Fig. \ref{babar_singlet}, we reinterpret the BABAR limits from Ref.~\cite{BaBar:2016sci} in the context 
of our scalar singlet model.\footnote{
Similar BABAR limits on muon-philic scalars were also calculated in Refs. \cite{Chen:2017awl} and \cite{Krnjaic:2019rsv}. 
Unlike Ref. \cite{Krnjaic:2019rsv}, which approximated the BABAR limit with a smooth curve, we 
apply the full mass dependence of the limit from the reported bounds from \cite{BaBar:2016sci}, which results in 
a more jagged curve in our Fig. \ref{babar_singlet} (right). Also in contrast to Ref. \cite{Krnjaic:2019rsv},
we apply the full geometric acceptance to the rescaling.
} Having already obtained the limit for vectors $V$, we can obtain the 
limit for scalars by applying the rescaling
\be
\label{rescaling}
g_S^{\text{limit}} = g_V^\text{limit}
\sqrt{
\frac{\epsilon_V}{\epsilon_S}
 \frac{\sigma(e^+e^- \to \mu^+ \mu^- V)}{\sigma(e^+e^- \to \mu^+ \mu^- S)}
\frac{\text{BR}(V\to \mu^+\mu^-) }{ \text{BR}(S\to \mu^+\mu^-) } 
}~,
\ee
where $\sigma$ is the singlet production cross
section, $\epsilon_{S/V}$ is the geometric 
acceptance, and each of the quantities here is mass-dependent. To obtain the necessary
quantities for this rescaling, we simulated $e^+e^-\to \mu^+\mu^- S/V$  production 
events using MadGraph5 \cite{Alwall:2014hca,Alwall:2011uj} which enable us to extract the cross sections and geometric acceptances assuming  $\theta \in [18^\degree ,150^\degree]$ angular coverage in the BABAR lab frame, where $\theta$ is the polar angle of the final-state muon
defined with respect to the incident electron beamline. 
We have also have verified that our simulated events
 adequately reproduce the $L_\mu -L_\tau$ limits reported in Ref.~\cite{BaBar:2016sci} when adapted to that scenario.

In both panels of Fig. \ref{babar_singlet} we also show future projections for a 4$\mu$ search at Belle II
with 50 ab$^{-1}$ of luminosity. Here we assume that the BABAR background model and statistical
uncertainties from Ref.~\cite{BaBar:2016sci}  can be adapted to the Belle II experimental setup with a rescaled luminosity. 
To place a limit on the singlet couplings, we demand that the signal
in each bin of di-muon reduced mass $m_{R} \equiv \sqrt{ m_{\mu^+\mu^-}^2 - 4m_\mu^2 }$ not exceed 
the expected SM background by more than 2$\sigma$, where $ m_{\mu^+\mu^-}$ is the reduced 
mass of an opposite state di-muon pair in the final state. A more rigorous, Belle II-specific 
systematic analysis is beyond the scope of this paper, but would be interesting for future work. 
Since the Belle II luminosity improves upon the BABAR data set by a factor of $\sim 100$, this 
search alone will cover all remaining scalar singlets with mass between $2m_\mu$ and 10 GeV unless their branching
fraction is to di-muons is less than $\sim$ 1\%.

\section{Collider Searches}
\label{s.colliders}

We now discuss direct singlet searches at the LHC in the 3- and 4-muon final states, as well as inclusive singlet searches at future muon colliders. Since LHC searches have limited mass reach for scalars, it is important to take electroweak precision constraints from all possible types of UV completions  for singlet scalar models into account, which sets a more restrictive upper mass limit than unitarity alone. Even so, we find that future muon colliders are required to exhaustively probe singlet explanations of the $\g$ anomaly.

\subsection{Singlet Production at the LHC}
\label{four-mu}

 \begin{figure}[t]
\center
\includegraphics[width=2.5in]{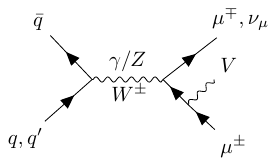} ~~~
\includegraphics[width=2.5in]{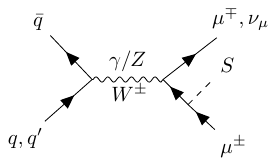}
\caption{Singlet production at a hadron collider via the neutral ($W$-mediated) and charged ($\gamma/Z$-mediated) Drell-Yan processes.
\label{f.lhc}}
\end{figure}

In order to produce $S$ or $V$ at a hadron collider, one needs to produce either one or two muons via charged $(p p\to \ell^\pm \nu_\ell)$ or neutral $(p p\to \ell^+ \ell^-)$ Drell-Yan processes and then let one of the muons radiate a singlet. These processes are depicted in Fig. \ref{f.lhc}. Because the singlets can decay promptly back to a muon pair, the overall reactions look like three- and four-muon production, respectively.
Backgrounds for these reactions include similar processes to those in Fig.~\ref{f.lhc}, replacing the singlets by a $\gamma/Z$ boson that further decays into a pair of muons, as well as $Z+Z/W$ production where the vector bosons decay leptonically. These irreducible backgrounds combined can be sizeable compared to the signals, but with proper cuts on the invariant mass of opposite-charge di-muon pairs, one can isolate the signals with great sensitivity in the case where the singlets decay 100\% to muons.

\begin{figure}[t!]
\includegraphics[width=2.8in]{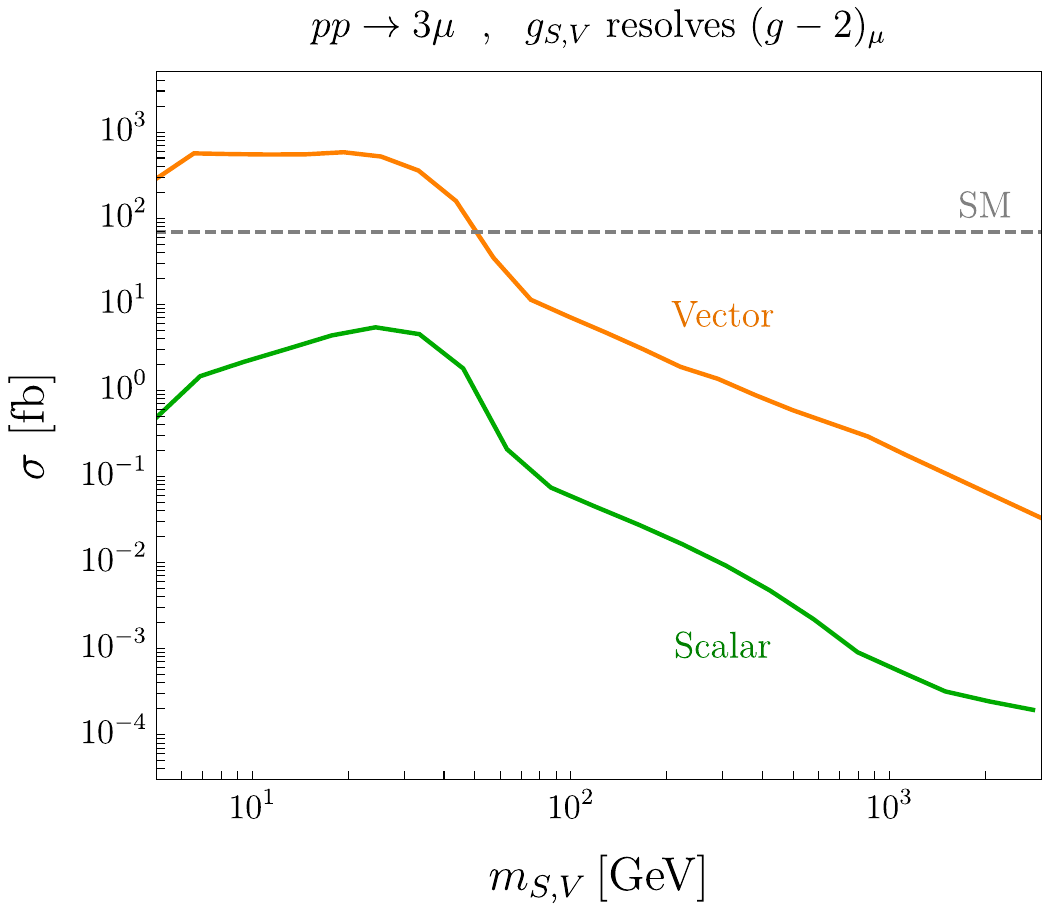}
\includegraphics[width=2.8in]{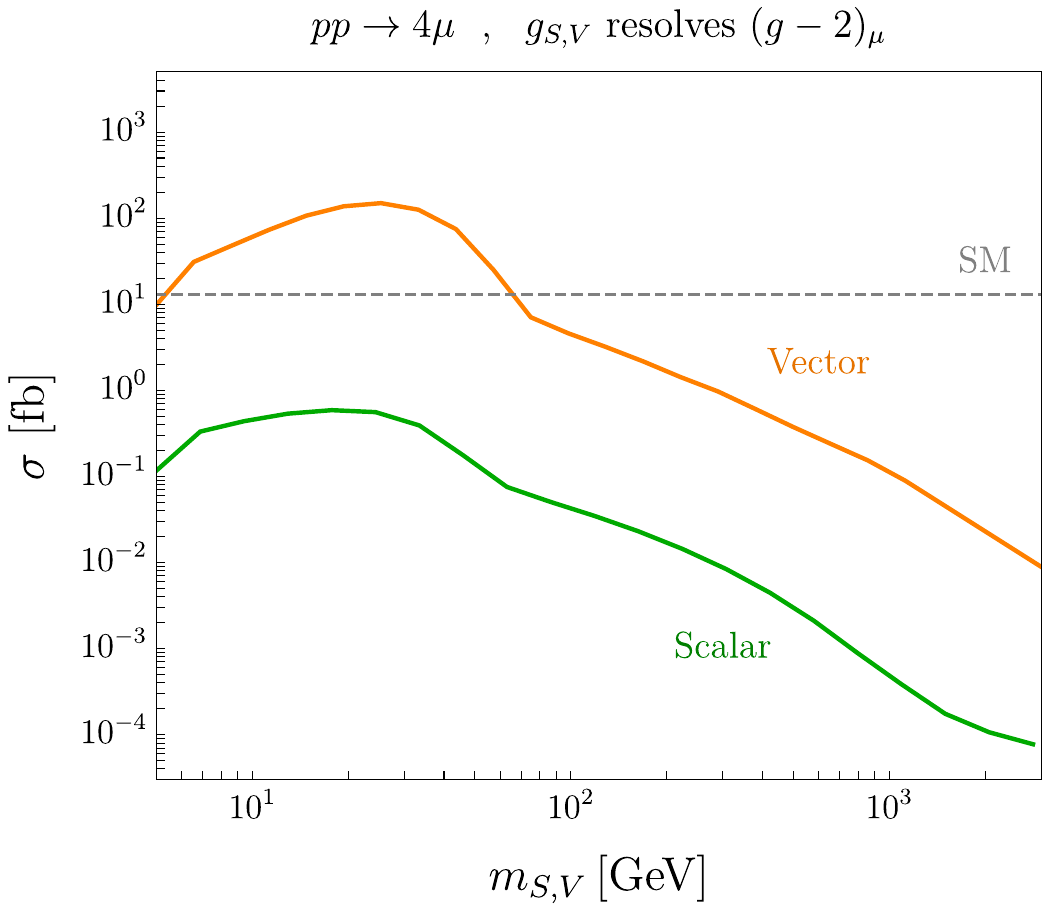}
\caption{  
LHC production cross sections for scalar and vector singlets in $pp \to 3\mu$ (left) and $pp \to 4\mu$ (right) final states that include $S/V \to \mu^+\mu^-$ contributions assuming 
$100\%$ branching ratios to di-muons. In all cases we assume $\sqrt{s} = 13$ TeV and the singlet couplings are chosen to resolve the $\g$ anomaly as shown in Fig. \ref{f.g-2contours}. 
The dashed gray line in each plot corresponds to the SM background prediction in each channel.
\label{f.xsec}}
\end{figure}

For our analysis, we implemented the Lagrangian \Eq{lag-singlets} in FeynRules2 \cite{Christensen:2008py,Alloul:2013bka} and generated charged and neutral Drell-Yan events with  MadGraph5 \cite{Alwall:2014hca,Alwall:2011uj}. (For this clean muon signal, detector resolution and efficiency effects play a subdominant role and can be neglected in our estimates.)
The cross sections for three- (left) and four-muon (right) production are presented in Fig.~\ref{f.xsec} where the SM background is shown in dashed gray and the vector and scalar singlets are shown in orange and green, respectively. These cross sections were obtained using the same loose cuts implemented in the four-lepton search by ATLAS in \cite{ATLAS:2021kog}: $p_{T}^{\mu} >20$ GeV (leading muon), $p_{T}^{\mu} >10$ GeV (sub-leading muon), $p_{T}^{\mu} >5$ GeV  (the other muons), $|\eta_{\mu}| <2.7$, $\Delta R_{\mu\mu} >0.05$, and $m_{\mu\mu} >5$ GeV, where the small angular acceptance is required to allow topologies where a boosted particle decays into a pair of muons, and the cut in the invariant mass $m_{\mu \mu}$ of any pair of opposite-sign muons is imposed to exclude leptons from quarkonia. These cross sections show that even without further cuts, the high-luminosity LHC (HL-LHC) should be able to probe singlet scalar masses up to $m_S \lesssim m_Z/2$, but as we show below, the mass reach can be significantly increased by making use of the resonant nature of on-shell singlet production.

Note that in Fig.~\ref{f.xsec}, the vector and the scalar production cross sections differ by about two orders of magnitude. This is for two reasons. First, the coupling $g_{S/V}$ fixed by $\g$ is larger for vector singlets than for scalars by about a factor of three in the region above $\sim 1$ GeV as shown in Fig. \ref{f.g-2contours} (left). Second, due to the chiral flip at the scalar vertex in \Eq{singlet-couplings}, the overall scalar production is a helicity-1 to a helicity-0 transition, so it is $p$-wave suppressed. This does not happen in vector models where an $s$-wave transition is possible, because the vector does not flip the chirality of the fermion lines. 
Also note that the charged Drell-Yan process ($3\mu$) has a larger cross section than the neutral one ($4\mu$). This is because $W$ couplings  to fermions are larger than $Z$ couplings, and because the $3\mu$ process includes two channels ($\mu^+\mu^-\mu^\pm$) whereas $4\mu$ includes only one channel ($\mu^+\mu^-\mu^+\mu^-$) and an extra suppression by statistical factors for identical particles in the final states.

Amongst recent LHC searches, the most relevant ones for this signal are two $139~\mathrm{fb}^{-1}$ analyses conducted by ATLAS searching for new physics in $3\mu$ and $4\mu$ final states~\cite{ATLAS:2021kog, ATLAS:2021wob}, as well as a $77~\mathrm{fb}^{-1}$ $Z\to 4\mu$ search for $L_\mu - L_\tau$ gauge bosons by CMS~\cite{CMS:2018yxg}.
The CMS search can be applied  almost verbatim to our singlet vector scenario. Assuming the acceptances do not change significantly, we can also recast this search for the singlet scalar using the ratio between the vector and scalar singlet $4\mu$ cross sections in \fref{f.xsec}. The resulting limits are shown in \fref{babar_singlet} and already exclude singlet scalar (vector) explanations of the $(g-2)_\mu$ anomaly in the mass range $m_S \sim 6 - 30 \gev$ ($m_V \sim 5 - 70 \gev$), if these singlets decay exclusively to muons.

The mass reach can be significantly extended by dropping the requirement of an intermediate on-shell $Z$ boson, and also considering $3\mu$ final states. This means that recent multi-muon ATLAS analyses~\cite{ATLAS:2021kog, ATLAS:2021wob} have great exclusion power for these singlet scenarios.
Unfortunately, these analyses do not explicitly look for an intermediate di-muon resonance. However, if  the di-muon invariant mass distribution we describe below were made public for this or future such analyses, then these searches could supply the best limits on singlet scenarios that resolve the $(g-2)_\mu$ anomaly.

To demonstrate this, we derive sensitivity projections by considering the combined distribution of invariant masses of all possible opposite-sign di-muon pairs  $m_{\mu^+\mu^-}$ in each event.
Because there are multiple muons in the final states, the combinatorial background tends to spread the invariant mass spectrum over a wide range of values. Even so, the singlet- and $Z$-resonances remain clearly visible in the signal and background. For each singlet mass, we define a signal region in this combined $m_{\mu^+\mu^-}$ distribution that is centered on $m_{S,V}$. The signal region has a variable width, ranging from 10 GeV for the lowest $m_{S,V}$ up to 30\% of the singlet mass for larger values of $m_{S,V}$.\footnote{This is highly conservative for narrow-width singlets. For singlets with decay width near the unitarity bound, our signal region as defined will not cover the full resonance width, but as our results indicate, for singlets with large invisible width, direct LHC searches for the muon final state are completely ineffective anyway.}

\begin{figure}[t]
\center
\hspace{-0.5cm}
\includegraphics[width=2.95in]{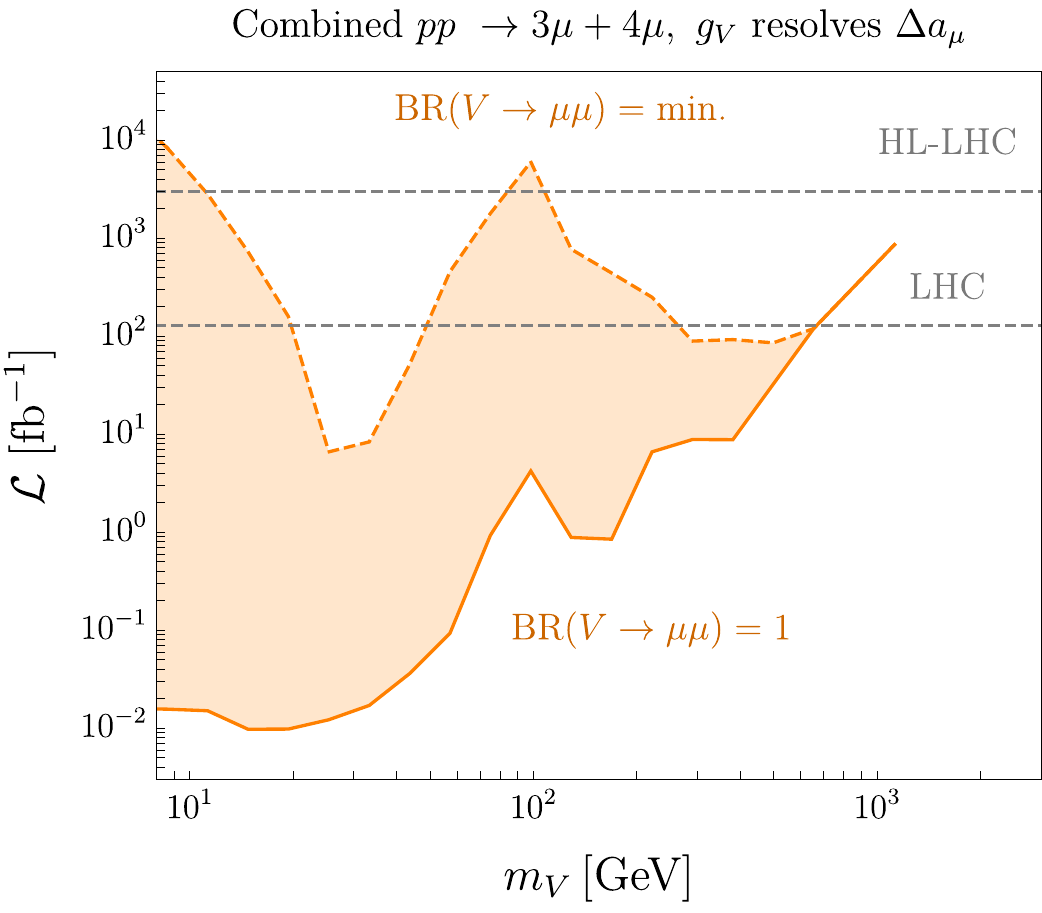}~~
\includegraphics[width=2.95in]{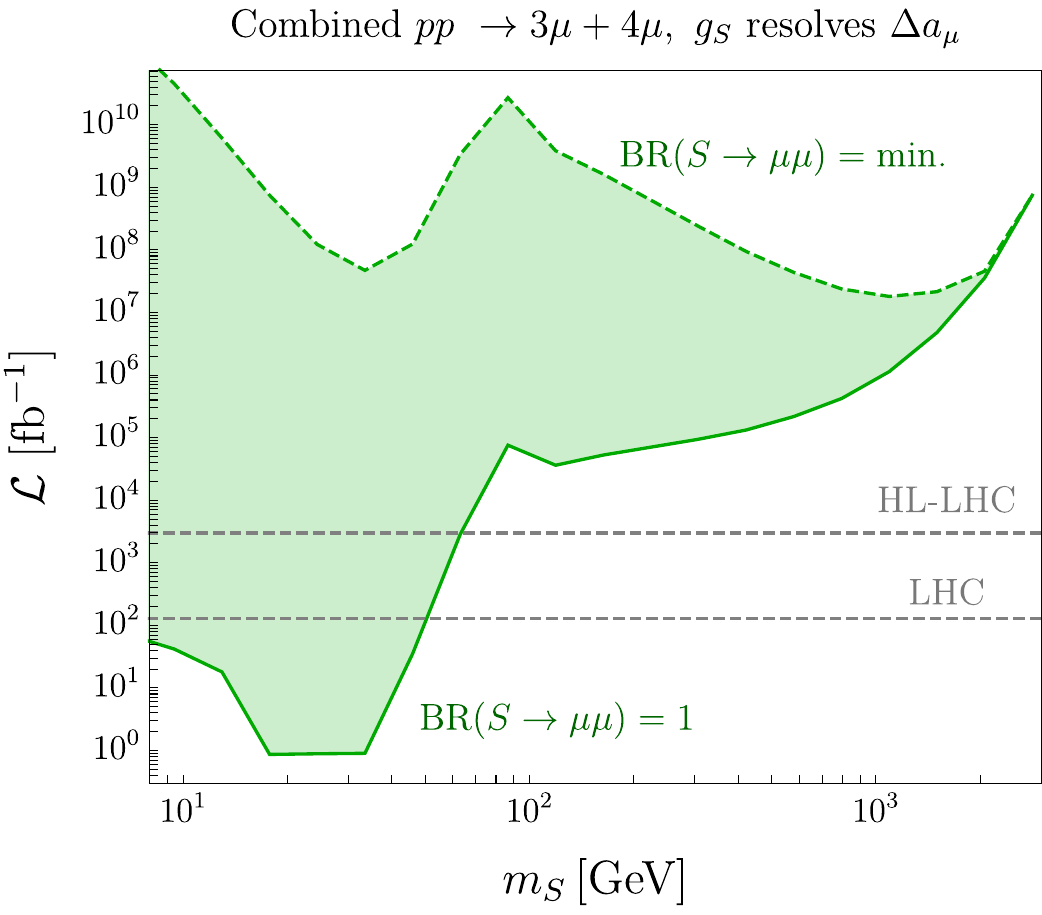}
\caption{
LHC luminosity required to exclude vector (left) and scalar (right) singlet models at 2 sigma via a resonant singlet search in the 3+4$\mu$ channel, for singlet couplings set to resolve the $(g-2)_\mu$ anomaly. The upper dashed curves corresponds to the minimum singlet branching fraction to muons, see Eqs.~(\ref{branch-min-vector}) and~(\ref{branch-min-scalar}). The lower solid curves correspond to singlets decaying entirely to muons. 
\label{f.bounds}}
\end{figure}

Fig. \ref{f.bounds}  shows the LHC luminosity required using this strategy to exclude singlet scenarios that resolve the $(g-2)_\mu$ anomaly for masses above 5 GeV.
This corresponds to a combined analysis of bump hunts in both charged and neutral Drell-Yan events, which we refer to from here on as a $3+4\mu$ search. The $3\mu$ events dominate the constraints for low singlet masses, whereas $4\mu$ events dominate for large masses. The solid curve corresponds to the optimistic scenario where the singlets decay with a 100\% branching ratio to muons, while the dashed curve corresponds to the minimum irreducible branching fraction to muons as shown in Fig. \ref{f.g-2contours} (right).

In the optimistic scenario, one can see that $3+4\mu$ searches at the LHC can probe singlet vector masses up to 700 GeV with current luminosity $140 \ {\rm ab}^{-1}$, whereas with the expected luminosity at HL-LHC the $3+4\mu$ search could probe the entire parameter space up to the perturbativity limit at about 1 TeV. As discussed above, the signal cross section for scalars is much smaller than for vectors for a given singlet mass. For this reason the reach of the $3+4\mu$ search is not as promising in the case of scalar singlets. With current (future) luminosity, the LHC could probe masses up to 50 (60) GeV. 
If the singlets only have the minimum irreducible branching ratio to muons, direct searches at the LHC have almost no power to probe this scenario. 
Even so, it is clear that this strategy greatly increases the mass reach compared to a search that is restricted to exotic $Z$ decays.
\fref{babar_singlet} shows these HL-LHC projections as bounds on $g_{S,V}$.

\subsection{Electroweak Precision Constraints and Singlet Scalar UV Completions}\label{EWPT}

As we can see from Fig.~\ref{f.bounds}, 
singlet scalars can only be directly probed with  $3+4\mu$ searches at the HL-LHC if $m_S \lesssim 60 \gev$. 
It seems very difficult to directly probe heavier  scalars coupling only to muons at a proton collider. Fortunately, we can exploit the fact that the scalar singlet model must have a UV completion featuring new electroweak charged states. 

\begin{figure}[t]
\center
\hspace{-0.5cm}
\includegraphics[width=4.6in]{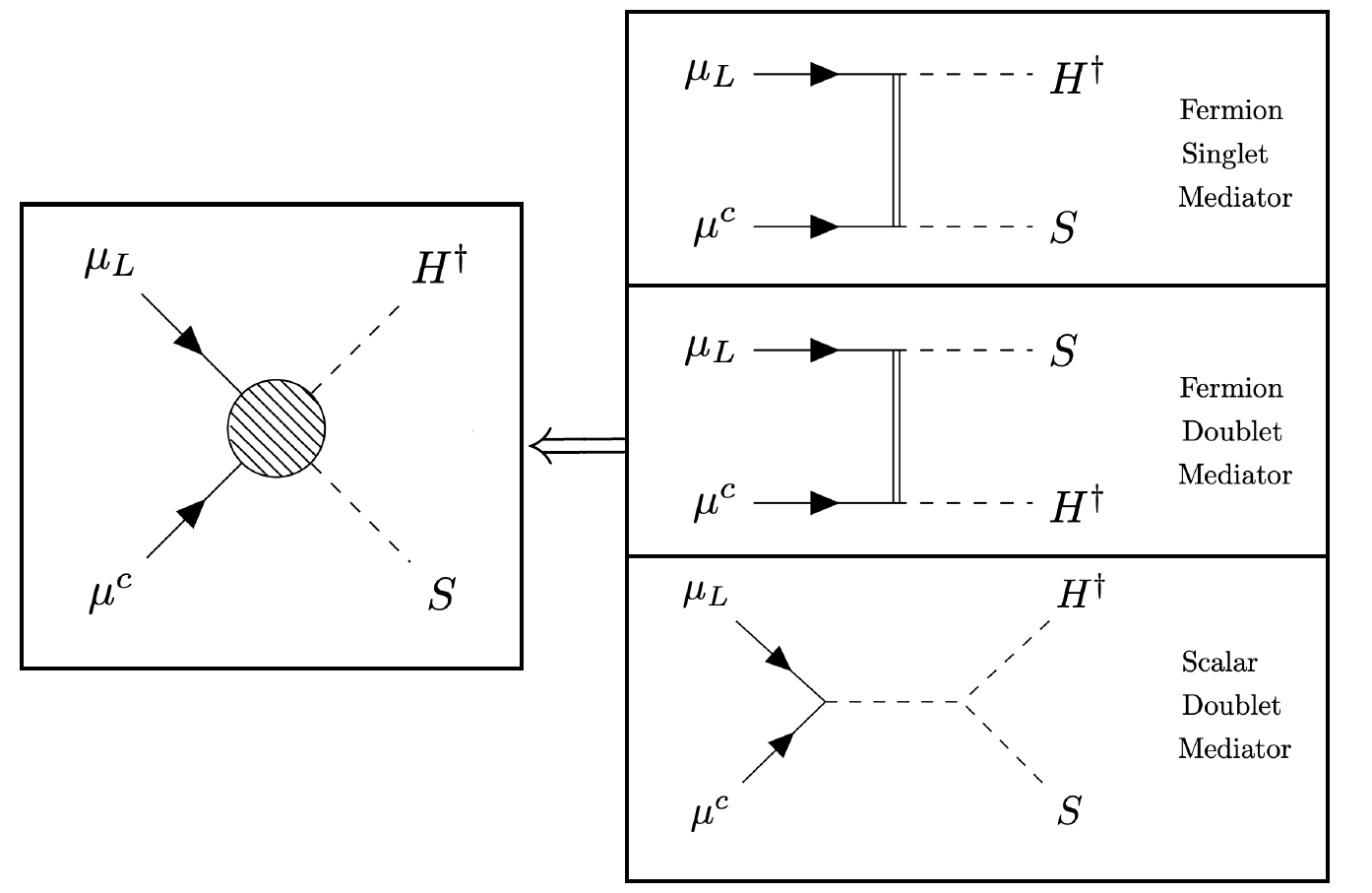}
\caption{
The singlet scalar scenario requires a UV completion to generate the dimension-5 operator $ \frac{1}{\Lambda} S H^\dagger L \mu^c$ (left), see \eref{singlet-yukawa-higher-dim}.
There are three possibilities to generate this operator via tree-level exchange of a mediator field, shown on the right. Assuming a minimal particle content for the additional fields, this corresponds to integrating out a fermion singlet ($t$-channel), a fermion doublet ($t$-channel), or a scalar doublet ($s$-channel).
}
\label{f.scalar_uv}
\end{figure}

As discussed in \sref{s.scalar_singlets}, the scalar Yukawa interaction $S \mu_L \mu^c$  in \Eq{lag-singlets} is not gauge-invariant and must arise from the dimension-5 operator
 \be
 {\cal L} \supset \frac{1}{\Lambda} S H^\dagger L \mu^c,
 \ee
where $\Lambda$ is some BSM mass scale.
We will assume that this operator is generated by tree-level exchange of some heavy mediator; if this interaction is somehow generated at higher loop level, the resulting electroweak states will be much lighter, resulting in even tighter constraints.
There are only three possible types of diagrams that can generate this operator at tree-level, shown in \fref{f.scalar_uv}: exchange of a fermion mediator that is an $SU(2)_L$ singlet, a  doublet fermion mediator, or a doublet scalar mediator. 
The top two diagrams to the right correspond to what we call {\it fermion UV completions} and the bottom-right diagram is the {\it scalar UV completion} for the singlet scalar model. After electroweak symmetry breaking, the former class features heavy fermions that mix with the muon, whereas the latter features a heavy scalar doublet that mixes with the scalar singlet $S$.

It is easy to understand why electroweak precision tests (EWPT)~\cite{ParticleDataGroup:2020ssz} can constrain these UV completions. If  the singlet accounts for the $(g-2)_\mu$ anomaly via an effective coupling to the muon $g_S \sim v/\Lambda$ as in \fref{f.g-2contours}, then heavier singlets require lower masses and/or larger couplings for the new mediator states. 
To derive indirect constraints on scalar singlet scenarios, we make the minimal assumption that the scenario is UV-completed by a single one of these three mediator types. 

The main EWPT we focus on in our analysis is the modified lepton universality ratio
\be
R_{\mu e}=\frac{\Gamma(Z\to\mu\mu)}{\Gamma(Z\to ee)}.
\label{rmue}
\ee
In the SM, this ratio is close to unity except for small phase space differences and small loop corrections. 
Under the assumption that the singlet scalar resolves the $(g-2)_\mu$ anomaly, both scalar and fermion UV completions generate significant $Z\mu\mu$ vertex corrections that can be probed by LEP measurements~\cite{ParticleDataGroup:2020ssz}. For scalar UV completions, the presence of additional scalars with electroweak charges also gives additional contributions to $3+4 \mu$ production at the LHC. 
The details of this analysis are provided in Appendix \ref{appendixB}. The important point is that we always consider the \emph{minimal} size of $R_{\mu e}$ allowed by adjusting the various mediator couplings within the bounds set by unitarity, and that the simple singlet scalar model of \Eq{singlet-couplings} is a good IR effective theory in the non-excluded regions of parameter space. Our main results are the following:
\begin{itemize}
\item Fermion UV completions of the singlet scalar violate EWPT for $m_S > 167\gev$. 
\item Scalar UV completions of the singlet scalar violate EWPT for $m_S > 615\gev$. 
\item If the singlet scalar only couples to the muons, then additional contributions to $3+4\mu$ production at the LHC, together with EWPT, exclude $m_S < 63 \gev$. 
\end{itemize}
Since these constraints were derived by directly relating the required size of the dimension-5 operator \eref{singlet-yukawa-higher-dim} (to generate $\Delta a_\mu$) to the corresponding \emph{minimal} $Z\mu\mu$ vertex correction, it is unlikely that these constraints would be greatly affected by considering more non-minimal UV completions, since additional couplings generally lead to additional observable effects unless deliberate cancellations are enforced. 
Therefore, it appears that with current LEP constraints, singlet scalar scenarios cannot be consistent with $S$ masses above roughly 615 GeV, which has implications for future colliders that could aim to directly produce these singlets. We indicate this upper mass bound as the vertical dashed gray line in \fref{babar_singlet}.

\subsection{Future Muon Colliders}
\label{MuC}

Muon colliders have recently attracted considerable
interest as successor machines to the LHC \cite{Alexahin:2018svu,Buttazzo:2018qqp,Boscolo:2018ytm,Neuffer:2018yof,DiLuzio:2018jwd,Delahaye:2019omf,Shiltsev:2019rfl,Costantini:2020stv,Han:2020uid,Han:2020pif,Han:2020uak,Bandyopadhyay:2020otm,Buttazzo:2020uzc,Huang:2021nkl,Liu:2021jyc,Chen:2021rnl,Han:2021udl,Collamati:20216A,Capdevilla:2021fmj,Huang:2021biu,Han:2021kes,Bottaro:2021srh,AlAli:2021let,Chiesa:2020awd,Long:2020wfp,Asadi:2021gah,Franceschini:2021aqd,Collamati:2021rbr,Sen:2021fha,Bandyopadhyay:2021pld,Qian:2021ihf,Chiesa:2021qpr,Liu:2021akf,Zaza:2021bj,Mastrapasqua:2021uvs,Zaza:2021nnq,Ruiz:2021tdt,Casarsa:2021rud}.
Representative muon colliders concepts at the energy frontier may
feature multi-TeV scale center-of-mass energies and 
luminosities of order $\sim$ $\rm{ab}^{-1}$. Such facilities
would be ideal laboratories for studying heavy, muon-philic interactions that 
cannot be probed robustly with existing proton or electron collider concepts \cite{Capdevilla:2021rwo}.
Indeed, it has been shown that such machines can perform model-independent tests 
of the $\g$ anomaly via muonic Bhabha scattering \cite{Capdevilla:2021rwo}
 and the $\mu^+\mu^- \to \gamma h$ channel  \cite{Buttazzo:2020ibd,Yin:2020afe}.

We have also shown in previous work that heavy singlets responsible for the $\g$ anomaly 
 can be directly tested through
 $\mu^+\mu^- \to \gamma S/V$ production independently of 
 how the singlets decay~\cite{Capdevilla:2021rwo}. Since 2-body particle production yields
 back-to-back final-state recoils, this strategy
triggers on events where the final-state photon
 is the only object in a detector hemisphere surrounding its momentum vector. Due to the large couplings required for heavy singlets (see Fig.~\ref{f.g-2contours}), such production events can exceed the SM prediction
 for hemispherically-isolated single photons. Furthermore, since the observable is the isolated
 photon, the sensitivity of this strategy is independent of how the singlet decays as it recoils away from the photon. In Fig. \ref{babar_singlet}
 we show muon collider projections for singlet sensitivity assuming a 215 GeV (3 TeV) center-of-mass energy
 and 0.4 ab$^{-1}$  (1 ab$^{-1}$) of integrated luminosity as purple solid (dashed) curves.
 For the 215 GeV muon collider, singlets below $\sim 100 \gev$ are probed via direct production, while heavier singlets are probed through
 their virtual contributions to muonic Bhabha scattering ($\mu^+\mu^- \to \mu^+\mu^-$). At a 3 TeV muon collider, direct production is always more sensitive as long as the singlet can be produced on-shell, so the analogous Bhabha projections do not contribute to the exclusion curve we show.

\section{Conclusions}
\label{s.conclusions}

In this paper we have systematically investigated solutions to $\g$ in which a SM singlet generates the dominant contribution to the observed discrepancy via renormalizable interactions with muons. There are only two such classes of models, involving a singlet scalar $S$ or a vector $V$, and both are highly restricted by general principles: scalars by the requirement of UV completion, and vectors by anomaly cancellation or non-decoupling triangle diagrams. Using only robust arguments from cosmology and unitarity, we constrain the viable mass range for singlets to lie between a few MeV and a few TeV~\cite{Capdevilla:2020qel,Capdevilla:2021rwo}. We have then surveyed a program of existing and future experiments to cover this parameter space, illustrated in \fref{babar_singlet} and summarized here:
\begin{itemize}

\item {\bf Model-Independent Probes:} \\
A future 215 GeV muon collider could cover the entire $\g$ favored region above $2 m_\mu$ for vectors and about 1 GeV for scalars. 
A 3 TeV muon collider is less sensitive to light singlets, excluding all favored masses above about 1 GeV for vectors and 10 GeV for scalars. 
Notably, this is independent of how the singlet decays, as it only relies on the photon kinematics in
$\mu^+\mu^- \to \gamma S/V$ events ~\cite{Capdevilla:2021rwo}, or singlet contributions to Bhabha scattering.
%

\item {\bf Di-muon Decays:}\\ If the singlet decays predominantly to di-muons (when kinematically allowed), the $\g$ parameter space can be
probed with a combination of $B$-factory and high-energy collider searches. Current BABAR and  LHC data, together with future Belle II and HL-LHC searches, can cover nearly the entire $\g$ parameter space for vectors above $2m_\mu$; scalar singlets will be robustly covered for masses between $2m_\mu$ and $60~\gev$.

\item {\bf Invisible Decays:} \\
Missing energy/momentum experiments (including NA64$\mu$, which will be taking preliminary data imminently) will cover the entire $\g$ region below $2m_\mu$, for both scalars and vectors which decay invisibly on the scale of the experiment, and have additional reach into the GeV scale if the dominant decay is invisible, as is true for viable anomaly-free vector models such as an $L_\mu - L_\tau$ gauge boson.

\item{\bf Other Visible Decays:} \\
The only remaining gap in the coverage presented here involves {\it prompt} visible decays to $e^+e^-, \gamma \gamma$, and $\pi^+\pi^-$ (or 
other visible exotics) for singlets in the $\sim$ MeV--10 GeV mass window; higher masses will be robustly covered with muon colliders, and
displaced decays will be covered by missing-energy/momentum experiments at which long-lived particles yield missing energy. However,
it is generically difficult to engineer prompt and visible decays for singlets that resolve $\g$ in this mass range. All anomaly-free vector models with visible decays in this range have already been ruled out \cite{Bauer:2018onh}, and anomalous vectors are likely subject to severe constraints from non-decoupling triangle diagrams. Scalar singlets can be viable, but the electron coupling is generically required to be smaller than the muon coupling, which would yield macroscopic decay lengths for much of the allowed mass range. Similarly, long decay lengths are expected
if the scalar decays through the minimal di-photon coupling induced by integrating out the muon \cite{Chen:2017awl}. We
leave a more detailed study of this remaining parameter space with non-muonic decay channels for future work.

\end{itemize}
Thus, a broad program of energy and intensity frontier probes can robustly cover all the currently viable $\g$ parameter 
space with only a few narrow exceptions. 

We have also studied how the singlet scalar model may be UV completed, and there are just three classes of possibilities if the singlet-muon coupling is generated by tree-level exchange of heavy mediators. Electroweak precision measurements then suggest that the singlet scalar mass has to be less than about 600 GeV, a factor of five lower than the upper bound provided by naive unitarity arguments alone.

Our analyses show that even without a muon collider, the parameter space of singlets that solve the $\g$ anomaly can be significantly probed by beam dumps, $B$-factories, and HL-LHC searches. However, this leaves large gaps in coverage. For the optimistic scenario where  the new singlet decays entirely to muons, scalars above 60 GeV are inaccessible. In the pessimistic scenario where the singlet only has the irreducible minimum branching ratio to muons required by unitarity, vectors between $\sim 20 - 200 \gev$ are inaccessible, while scalars represent something of a nightmare scenario, apparently impossible to see above $\sim 1 \gev$ with this suite of experimental probes.

On the other hand, it is clear that a muon collider is required to probe singlet scalar solutions to the $\g$ anomaly. They are straightforward to UV complete with masses much larger than the HL-LHC reach, and it is similarly easy to imagine scenarios where they have other decays that hide them from muon searches. 
Furthermore, probing singlet solutions particularly motivates a staggered approach in center-of-mass energy for the muon collider program. A 3 TeV machine is obviously preferred to push the boundaries of the energy frontier, as well as to start probing solutions to $\g$ that involve dominant contributions by new electroweak charged states~\cite{Capdevilla:2021rwo}.
However, such a high-energy machine actually has more difficulty discovering light singlet scalars below 10 GeV, while direct production at a 215 GeV muon collider is sensitive down to the upper mass limit of proposed muon fixed-target experiments. 

It is interesting to consider whether other experimental probes could fill the gap up to scalar masses of 10 GeV, so that no gap in coverage remains  if only a multi-TeV muon collider were built. 
For both the scalar and the vector, the limited number of visible final states available in this narrow mass range should allow a complete enumeration of all of the possible visible and invisible decays for total widths approaching the unitarity limit. The outcome of such a study could have major implications for the future muon collider program by setting the minimum mass scale of new physics that such a high-energy machine would have to probe in order to leave no gaps in coverage. We leave this for future investigation.

The $\g$ anomaly is quite possibly our best hint of new BSM physics in the laboratory. Our work highlights the need for muon colliders to comprehensively probe the new physics that must exist if the anomaly is real, from very high masses~\cite{Capdevilla:2020qel,Capdevilla:2021rwo} all the way down to the GeV scale. No other kind of experiment has such a wide dynamic range in energy and precision, and if SM explanations of the anomaly are ultimately excluded, a comprehensive muon collider program would be well motivated as a top priority for the international particle physics community.

\vspace{5mm}
\textbf{Acknowledgements:} 
We thank  Nikita Blinov, Brian Shuve, Nathaniel Craig, and Kohsaku Tobioka 
for helpful conversations.
The research of RC and DC was supported in part by a Discovery Grant
from the Natural Sciences and Engineering Research Council of Canada,  the Canada
Research Chair program, the Alfred P. Sloan Foundation, and the Ontario Early Researcher Award. 
The work of RC was supported in part by the Perimeter Institute for Theoretical Physics (PI). Research at PI is supported in part by
the Government of Canada through the Department of
Innovation, Science and Economic Development Canada
and by the Province of Ontario through the Ministry of
Colleges and Universities.
The work of YK was supported in part by US Department of Energy grant DE-SC0015655. 
 This manuscript has been authored by Fermi Research Alliance, LLC under Contract No. DE-AC02-07CH11359
with the U.S. Department of Energy, Office of High Energy Physics.

\begin{appendix}

\section{Appendix:   Two-Loop Models}\label{appendixA}

Throughout this paper, we have studied singlet particles that couple 
linearly to muon currents and resolve the $\g$ anomaly at one loop through the Feynman
 diagrams shown in Fig. \ref{f.feynmang-2}.
 However, it is logically possible that the leading contribution to $a_\mu$ involves two-loop diagrams instead.
 In this appendix we assess the feasibility of 
 two representative
 two-loop scenarios.\footnote{We thank Nathaniel Craig for posing this challenge.} 
 
 \subsection{Bilinear Scalar Couplings}

In principle, a quadratic (or higher polynomial) scalar coupling
could yield appreciable contributions to $\g$ involving two-loop diagrams with no corresponding one-loop
contributions. 
By SM gauge invariance, the lowest-dimension operator that yields a two-loop correction to $\g$ involves a scalar $S$ 
with the interaction 
\be
\label{Z2}
{\cal L}_{\rm int} =  C''_\mathrm{eff} \frac{v}{\Lambda^2} S^2 \mu_L \mu^c + {\rm h.c.} ~,
\ee
where $\Lambda$ is a mass scale that arises from integrating out particles charged under SU(2)$_L \times$ U(1)$_Y$
and we assume that the simpler $S \mu_L \mu^c$ Yukawa coupling is negligibly suppressed or forbidden due to a  $\mathbb{Z}_2$ or other symmetry under which the $S$ is charged.

By naive dimensional analysis, the contribution to $\g$ from the  interaction in \Eq{Z2} is estimated
to be 
\be
\label{del_a}
a_\mu^{\rm BSM} \sim \brac{1}{16 \pi^2}^2 \left( C''_\mathrm{eff} \frac{v}{\Lambda^2} \right)^2 \frac{m_\mu^6}{ m_S^4}
\sim 5\times 10^{-10} (C''_\mathrm{eff})^2  \brac{90 \rm\, GeV}{\Lambda}^4
\brac{100 \, \rm MeV}{m_S}^4,
\ee
where the prefactor $(16 \pi^2)^{-2}$ arises from the two-loop phase space and we demand $\Lambda \gtrsim 90$ GeV to avoid model-independent  LEP bounds on new electroweak charged particles \cite{Egana-Ugrinovic:2018roi} while maximizing their contribution to $\g$.     
Furthermore, this correction to $\g$ saturates to a fixed value for $m_S \ll m_\mu$,  so the normalization in \Eq{del_a} 
represents the {\it maximum} contribution to the anomaly due to this interaction.
Despite these extremal choices, the magnitude of the effect in \Eq{del_a} appears barely large enough to account for 
the necessary $\Delta a_\mu \approx 2.5 \times 10^{-9}$ from \Eq{amu-exp} even if $C''_\mathrm{eff}$ is sizeable, but is sufficiently close  that a more careful calculation would be needed before fully excluding this possibility.

However, the interaction in \Eq{Z2} assumes that the two-loop interaction arises from integrating out a {\it single} particle 
with a coupling-to-mass ratio of order $\Lambda$, without introducing any intermediate steps. In practice, resolving  
 this higher-dimension operator with renormalizable and gauge-invariant interactions at energies above the scale $\Lambda$ is more challenging and further suppresses the BSM contribution to $\g$, effectively resulting in $C''_\mathrm{eff} \ll 1$.  Indeed, the  only  single-particle 
 UV completion of \Eq{Z2} involves coupling $S$ to the Higgs doublet $H$ and integrating out the Higgs boson after EWSB to obtain
 \be
 {\cal L} = \lambda_{HS} H^\dagger H S^2  + y_\mu H^\dagger \mu_L \mu^c + {\rm h.c.} \to \frac{ \lambda_{HS} v y_\mu  }{m_h^2} S^2 \mu_L \mu^c + {\rm h.c.}~,
 \ee
where $m_h = 125$ GeV is the Higgs mass, $y_\mu \sim 10^{-4}$ is the SM muon Yukawa coupling, and $\kappa$ is a dimensionless parameter. Here $\Lambda \to m_h$ and $C''_\mathrm{eff} \to \kappa y_\mu$, and for any unitary choice of $\lambda_{HS}$, the $a_\mu^{\rm BSM}$ contribution of \eref{del_a} is much too small to explain the anomaly.

Avoiding this Yukawa suppression requires adding additional SM charged states that couple to both the muon and the Higgs (for example, chiral fermions which mass-mix with the muon). These states must also couple to the SM-singlet operator $S^2$, which is dimension-2 and cannot couple to a dimension-3 fermion bilinear with a renormalizable interaction, so an additional scalar must also be added and then integrated out to generate $S^2 \mu_L \mu^c$ at low energies. However, if this additional scalar couples linearly to the $\mu_L \mu^c$ bilinear, it introduces quantitatively larger one-loop contributions to $\g$, which invalidates our motivation; if its coupling to this bilinear arises from some other interaction, even more fields must be added to the theory and integrated out.  
 Furthermore, despite these difficulties, the net effect of performing all these steps must still yield $C''_\mathrm{eff} v^2/\Lambda^2 \sim {\cal O}(1)$ in  \Eq{Z2}, which seems implausible. While we have not presented a rigorous theorem to eliminate such a possibility, even if it were possible to generate a large prefactor and suppress all one-loop corrections to $\g$, its contribution would still be too small based on \Eq{del_a}, subject to  the same caveats mentioned above.

\subsection{Millicharged Particles}
It has also been argued that a two-loop contribution to $\g$ could arise from a large $N_\chi \gg 1$ multiplicity of
 MeV-scale fermions $\chi$ with electromagnetic millicharges $\epsilon \ll 1$ \cite{Bai:2021nai}. In this scenario, loops of $\chi$ introduce new QED-like vacuum polarization diagrams whose contribution resolves the  anomaly if $N_\chi \epsilon^2 \sim 10^{-3}$.  
 
 However, such particles are subject to stringent limits from the E137 electron beam-dump experiment \cite{Bjorken:1988as,Batell:2014mga}, which were not considered in Ref. \cite{Bai:2021nai}. 
  At E137, millicharged particles could be radiatively produced in the beam dump via electron-nucleus scattering
   $e^- N \to  e^- N \chi \bar\chi$, pass unimpeded through the beam dump, and scatter in the downstream detector. Since E137
   reported null results, there are nontrivial limits on similar models involving MeV-scale scalars, as studied 
   in Ref. \cite{Batell:2014mga}. It is expected that this limit would also impose nontrivial bounds on 
   the millicharge model that explains $\g$ with two-loop diagrams. Such an analysis is beyond the scope of this 
   paper, but is worth studying in future work. 
 
 
 \section{Appendix:   $B$-factory Searches for $S/V \to e^+e^-, \gamma\gamma$}
\label{Bfactoryappendix}
For $m_{S,V} < 2 m_\mu$ it may be possible to perform a $B$-factory search for $S \to \gamma \gamma$ and $S/V \to e^+e^-$ decays, but understanding the relevant SM backgrounds calls for a dedicated analysis, which is beyond the scope of
this paper. Since the singlet coupling to electrons is, in principle, unrelated to the muon coupling 
that resolves the $\g$ anomaly, the decays can be appreciably displaced at $B$-factory energies. For example, if the singlet decays dominantly to electrons, its decay length is approximately
\be
L_S \approx 4.2 \text{ m}  \brac{E_S}{5 \, \rm GeV}\brac{10^{-4}}{g_{S,e}}^2 \brac{10 \, \rm MeV}{m_S},
\ee
where we have used the rest frame width $\Gamma(S \to e^+e^-) = g_{S,e}^2 m_S/(8\pi)$ and $g_{S,e}$ is the singlet coupling to electrons. 
Thus, a major challenge of this search strategy is distinguishing
the displaced decay signal from SM photon conversion backgrounds in which an $e^+e^-$ pair
is produced in photon-nucleus interactions in the detector. Similar considerations apply to 
$S \to \gamma \gamma$ decays which proceed through a higher-dimension operator (see Sec.~\ref{beam-dump})
and are expected to be similarly long-lived if the \emph{only} couplings in the model are the muon coupling and the loop-induced $\gamma \gamma$ coupling. We emphasize again that the singlet lifetime is unrelated to the singlet-muon coupling and can vary considerably, so it is not currently known whether a direct search strategy is possible for these specific final states. 

In the context
of discovering singlets responsible for $\g$, these visible non-muonic decay searches 
require $m_S < 2m_\mu$ where it is still possible for singlets to decay to $e^+e^-$ or $\gamma \gamma$ with a large branching fraction; 
viable singlets above the di-muon threshold will always have a larger branching fraction to di-muons\footnote{Various searches
for singlets decaying to $e^+e^-$ have excluded the $\g$ parameter space for particles
that couple with equal strength to muons and electrons \cite{Bauer:2018onh}. While
the analyses in Ref. \cite{Bauer:2018onh} emphasize vector particles, the bounds
on equal muon/electron coupling models greatly exceed the parameter space favored by $\g$ and, therefore, also apply
to scalar singlets whose signal strength only differs by order-one amounts.} or to invisible final states, in which
case the $4\mu$ search described in Sec.~\ref{s.bfac} or the invisible missing energy/momentum searches from Sec. \ref{missing-energy/momentum} are better strategies, respectively. Furthermore, singlets with very displaced decays to $e^+e^-, \gamma \gamma$  final states 
can also be discovered or falsified at missing energy/momentum 
experiments if they decay downstream of the detector to fake a missing energy signature (see Sec. \ref{missing-energy/momentum}).
Thus, the $B$-factory searches for $e^+e^-$ or 
$\gamma \gamma$ final states
 described here are only sensible for singlets that decay promptly to these particles 
 in the $m_S < 2m_\mu$ regime.

 
\section{Appendix:   Scalar Singlet UV Completions}\label{appendixB}

As discussed in Sec. \ref{EWPT}, we will investigate the constraints on singlet scalars in the context of the fermion and scalar UV completions illustrated in Fig. \ref{f.scalar_uv}. Some of these models were studied in detail in \cite{Bissmann:2020lge}. The Lagrangians for these interactions are
\begin{align}
\label{e.uv_lag}
\mathcal{L}_{I}   & \supset-y_{1}LH^{\dagger}\chi^{c}-y_{2}\mu^{c}\chi S + m_\chi \chi^c \chi, \nonumber \\
\mathcal{L}_{II}  & \supset-y_{1}L\Psi^{c}S-y_{2}\mu^{c}H^{\dagger}\Psi + m_\Psi \Psi^c \Psi, \\
\mathcal{L}_{III} & \supset-yL\Phi^{\dagger}\mu^{c}-\kappa SH^{\dagger}\Phi + m_\Phi^2 \Phi^* \Phi, \nonumber
\end{align}
for the fermion singlet mediator,  fermion doublet mediator and scalar doublet mediator respectively (for simplicity we are suppressing the ``$+ {\rm h.c.}$'' in all Lagrangians in this section). Here  $\chi\equiv({\textbf 1},-1)$ and $\Psi\equiv({\textbf 2},-1/2)$ are the new fermions, $\chi^c$ and $\Psi^c$ their corresponding vector-like partners, and $\Phi\equiv({\textbf 2},1/2)$ is the scalar doublet.

\subsection{Fermion UV Completions}

The first two Lagrangians above can be expanded into
\begin{align}
\mathcal{L}_{I}   & \supset -y_1\mu_L \chi^c H^{*} - y_2\mu^c\chi S, \nonumber \\
                          & \supset -y_1\bar{\mu} P_R \chi H - y_2 \bar{\mu} P_L \chi S, \\
\mathcal{L}_{II}  & \supset -y_1\mu_L \psi^c_d S - y_2 \mu^c \psi_d H^{*}, \nonumber \\
                          & \supset -y_1 \bar{\psi}_d P_L \mu S - y_2 \bar{\psi}_d P_R \mu H,
\end{align}
where the second lines are written in four-component spinors. After integrating out $\chi$ and $\psi_d$ (the down component of the doublet $\Psi$) one generates the dimension-5 operator
\be
 {\cal O} \supset \frac{y_1 y_2}{M} S H^{*} \mu_L \mu^c~,
\ee
where $M$ represents either the mass of $\chi$ or $\psi_d$. Once the Higgs gets a VEV, the scalar singlet Yukawa interaction $ {\cal L} \supset g_S S \mu_L \mu^c$ is generated, and we can identify the $\g$ coupling as
\be
 g_S = \frac{y_1 y_2 v }{\sqrt{2}M}.
\ee

\subsubsection*{Fermion Singlet UV Completion}

The mass matrix for $\mathcal{L}_{I}$ is given by
\be
{\cal L}_{{\rm mass}}\supset(\mu_{L}\,\chi)\left(\begin{array}{cc}
\tilde{y}_{e} & \tilde{y}_{1}\\
0 & m_\chi
\end{array}\right)\left(\begin{array}{c}
\mu^{c}\\
\chi^{c}
\end{array}\right),
\ee
where $\tilde{y}_{e}=y_{e}v /\sqrt{2}$ and $\tilde{y}_{1}=y_{1}v /\sqrt{2}$. We can diagonalize this matrix with two real rotations $M_{\rm diagonal} = L^T MR $, where
\be
L/R=\left(\begin{array}{cc}
\cos \theta_{L/R} & \sin \theta_{L/R}\\
-\sin \theta_{L/R} & \cos \theta_{L/R}
\end{array}\right)
\ee
and we can expand to obtain
\be
&& \cos \theta_{L}\sim1-\frac{1}{2}\left(\frac{\tilde{y}_{1}}{m_\chi}\right)^{2}~~,~~~  \cos \theta_{R}\sim1,\\
&& \sin \theta_{L}\sim\frac{\tilde{y}_{1}}{m_\chi}~~,~~~~~~~~~~~~~~~~~~  \sin \theta_{R}\sim\frac{\tilde{y}_{1}}{m_\chi}\frac{\tilde{y}_{e}}{m_\chi}.
\ee
The couplings between the $Z$ boson and the muon are
\be
\mathcal{L} \supset (g_L \mu_L^\dagger \bar\sigma_\nu \mu_L + g_R \mu^c \sigma_\nu \mu^{c \dagger}) Z^\nu.
\ee
Due to the mixing with $\chi$ the coupling $g_L$ gets shifted:
\be
g_L\to g_L+\sin^2 \theta_L (g_R-g_L).
\ee
This shift is what modifies the ratio $R_{\mu e}$ defined in Eq.~(\ref{rmue}) with respect to its SM value.

\subsubsection*{Fermion Doublet UV Completion}

The mass matrix for $\mathcal{L}_{II}$ is given by
\be
{\cal L}_{{\rm mass}}\supset(\mu_{L}\,\chi)\left(\begin{array}{cc}
\tilde{y}_{e} & 0\\
\tilde{y}_{2} & m_\psi
\end{array}\right)\left(\begin{array}{c}
\mu^{c}\\
\chi^{c}
\end{array}\right),
\ee
where this time the rotations are given by
\be
 && \cos \theta_{L}\sim1~~,~~ ~~~~~~~~\cos \theta_{R}\sim1-\frac{1}{2}\left(\frac{\tilde{y}_{2}}{m_\psi}\right)^{2},\\
 && \sin \theta_{L}\sim\frac{\tilde{y}_{2}}{m_\psi}\frac{\tilde{y}_{e}}{m_\psi}~~,~~ \sin \theta_{R}\sim\frac{\tilde{y}_{2}}{m_\psi}.
\ee
In this case the right-handed $Z\mu\mu$ coupling gets shifted:
\be
g_R\to g_R-\sin^2 \theta_R(g_R-g_L).
\ee

\begin{figure}[t]
\center
\hspace{-0.5cm}
\includegraphics[width=4in]{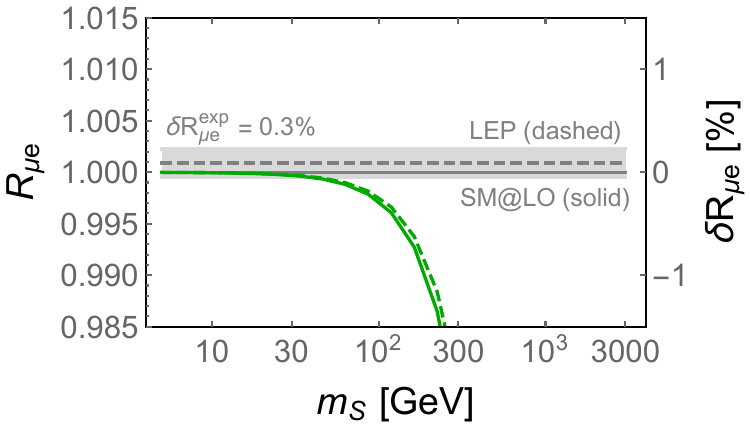}
\caption{SM expectation for the ratio $R_{\mu e}$ and its \emph{minimum} deviation due to muon mixing if a singlet scalar model UV completed by fermion mediators resolves the $(g-2)\mu$ anomaly. The gray solid line represents the leading order-calculation, and the gray dashed line with shaded band shows the experimental result from LEP with an error of $0.3\%$~\cite{ParticleDataGroup:2020ssz}. The solid (dashed) green line represents the fermion UV completion $\mathcal{L}_I$ ($\mathcal{L}_{II}$) from Eq.~(\ref{e.uv_lag}).}
\label{f.Rmue}
\end{figure}

\subsubsection*{Constraints}

Fig.~\ref{f.Rmue} shows the SM expectation for the ratio $R_{\mu e}$ and the corresponding \emph{minimum} deviation from the fermion UV completions $\mathcal{L}_I$ (solid) and $\mathcal{L}_{II}$ (dashed). 
These constraints were derived in the following way. For the $\mathcal{L}_I$  UV completion, the relevant mixing angle of the new fermion with the muon is $\theta \sim y_1v/M$, while
 $g_S = \frac{y_1 y_2 v }{\sqrt{2} M}$.
 To resolve the $(g-2)_\mu$ anomaly, $g_S = g_S(m_S)$ is fixed as a function of the singlet mass, see \fref{f.g-2contours} (left).
Therefore, for a given $m_S$, the mixing angle is fully determined $\theta \sim g_S(m_S)/y_2$, and is minimized by choosing $y_2$ to be as large as possible, in our case at the unitarity limit of $\sqrt{4 \pi}$. For $\mathcal{L}_{II}$ the argument is identical up to $y_1 \leftrightarrow y_2$. The green lines in Fig.~\ref{f.Rmue} translate into $2\sigma$ deviations in the ratio $R_{\mu e}$ for singlet masses $m_S$ above 167 GeV for the fermion UV completion $\mathcal{L}_I$, and above 181 GeV for the fermion UV completion $\mathcal{L}_{II}$.

\subsection{Scalar UV Completion}

\begin{figure}[t!]
\center
\hspace{-0.5cm}
\includegraphics[width=2.8in]{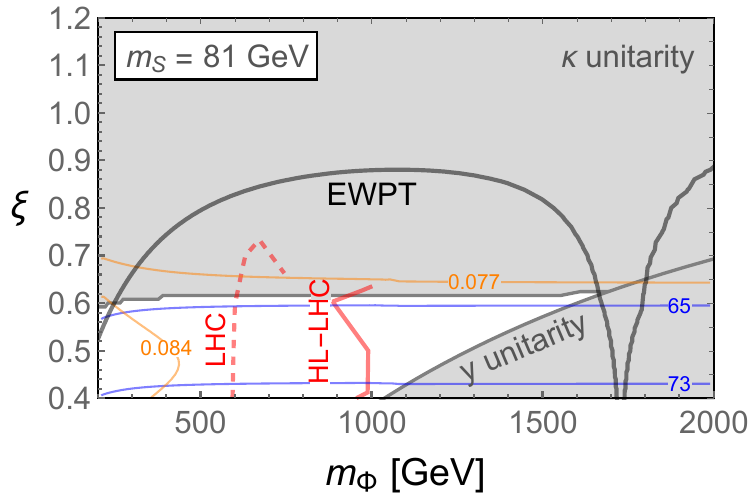} \includegraphics[width=2.8in]{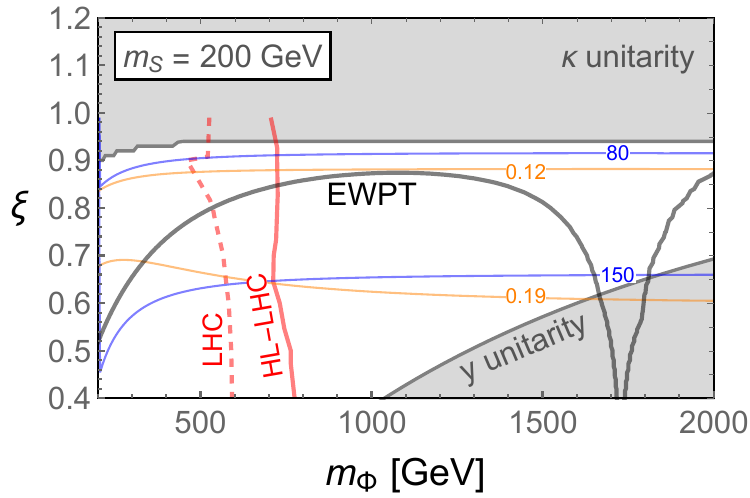} \\
\includegraphics[width=2.8in]{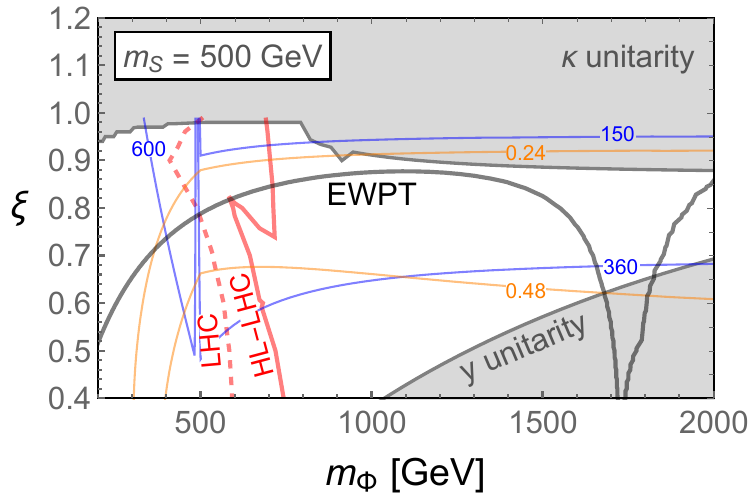} \includegraphics[width=2.8in]{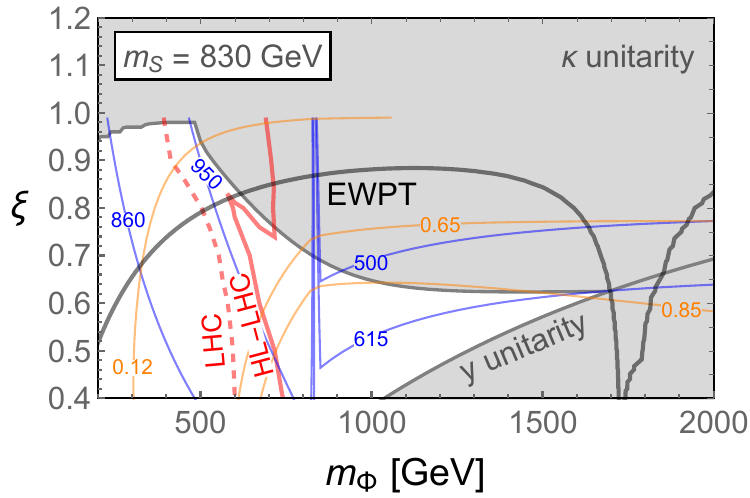}~~~
\caption{
Constraints on the parameter space of the scalar UV completion for the singlet scalar model. \emph{Note that $m_S, m_\Phi$ are now mass parameters in $\mathcal{L}_{III}$ (Eq.~(\ref{e.uv_lag})), not mass eigenvalues.} The parameter $\xi = \kappa v/m_S m_\Phi $ controls the mixing between the singlet $S$ and the scalar doublet mediator $\Phi$. At each point in this parameter space, the coupling $y$ is chosen so that $a_\mu^{\rm BSM} = \Delta a_\mu$. The bottom-right region requires $y$ couplings that violate unitarity. For large values of $\xi$, the trilinear coupling $\kappa$ violates perturbative unitarity. 
The region below the black solid line is excluded by EWPT as described in the text. 
The regions to the left of the dashed (solid) red lines would be excluded by a conservative search in the 3+4$\mu$ channels at the (HL-)LHC, see text.
The blue (orange) contours represent the mass of the $\varphi_1$ eigenstate that is mostly the singlet $S$ (the coupling of this eigenstate to muons), corresponding to the parameters $m_S$ ($g_S$) in the singlet scalar effective theory~(\ref{singlet-couplings}).
}
\label{f.ScalarUV}
\end{figure}

The third Lagrangian in Eq.~(\ref{e.uv_lag}) can be expanded as follows:
\be
\mathcal{L}_{III} \supset -y(\nu \phi_u^* + \mu_L \phi_d^*)\mu^c - \kappa H S \phi_d,
\label{alone}
\ee
where $\phi_{u,d}$ are the up and down components of the scalar doublet $\Phi$. The parameter space of this UV completion is defined by the mass of the singlet $m_S$, the mass of the doublet $m_\Phi$, the trilinear parameter $\kappa$, and the doublet-muon coupling $y$. For simplicity, we trade $\kappa$ for the mixing parameter
\be
\kappa = \frac{\xi m_S m_\Phi}{v},
\ee
where $\xi=0$ represents no mixing and $\xi=1$ represents maximal mixing so that one of the eigenstates in the theory becomes massless. (We do not consider values of $\xi>1$ for which the new scalars acquire VEVs and contribute to electroweak symmetry breaking.) For a given choice of $m_S$ (the mass parameter, not the mass eigenvalue) one can scan the plane $(\xi, m_\Phi)$ by fixing the coupling $y$ so that $ a_\mu^{\rm BSM} =\Delta a_\mu$ at each point in that plane.

The summary of constraints on the parameter space of the scalar doublet UV completion is shown in Fig.~\ref{f.ScalarUV}. Grey regions at the bottom and top of the figures are excluded by perturbative unitarity in the $y$ and $\kappa$ couplings respectively. The region below the solid black line is excluded by EWPT.
 After mixing, this model contains four scalar eigenstates $\varphi_1$ (mostly $S$), $\varphi_2$ (mostly $\phi_d$ CP-even), $\varphi_3$ ($\phi_d$ CP-odd), and $\phi_u$ (the charged component of $\Phi$). The blue contours represent the mass of the $\varphi_1$ eigenstate and the orange contours show the coupling of this eigenstate to muons, corresponding to $m_S$ and $g_S$ in the singlet scalar effective theory~(\ref{singlet-couplings}).
 The regions to the left of the red lines would be excluded from $3+4\mu$ searches at the LHC with current luminosity (dashed) and high luminosity (solid), which we discuss in more detail below.

The plots in Fig.~\ref{f.ScalarUV} show that for $m_S<81$ GeV and $m_S>830$ GeV, the multiple constraints do not leave any viable parameter space for the model to generate $\g$, provided the new scalars do not have additional couplings to hidden sector states that reduce the branching fraction to muons. This is the reason why the scalar UV completion is only valid in the interval $m_1\in(63, 615)$ GeV, where $m_1$ is the mass of the eigenstate $\varphi_1$. (Note the blue contours in Fig. \ref{f.ScalarUV} for the $m_S<81$ GeV and $m_S>830$ GeV panels.) If the singlet has additional couplings to invisible states, then the 3+4$\mu$ constraints would be weaker. We do not study this in detail, and therefore conservatively drop the lower bound on $m_1$ in this case. The upper bound is set by EWPT and is unaffected.

We now provide more details on each of the constraints shown in Fig. \ref{f.ScalarUV}.

\begin{itemize}

  \item \textbf{Perturbative unitarity in $y$:} In principle, the $y$ interaction in Eq.~(\ref{alone}) can produce the required contribution to $\g$ by itself. This interaction requires a non-perturbative $y$ coupling for masses above 560 GeV. By increasing the mixing between $S$ and $\phi_d$ via increasing $\xi$, one reduces the contribution to $\g$ from the doublet and increases the contribution from the singlet $S$. This means that the larger the mixing, the larger the $m_\Phi$ allowed by perturbativity of the coupling $y$. This explains the ascending behavior of the ``$y$ unitarity'' constraint in the plots.
  
  \item \textbf{EWPT:} When the mixing $\xi$ is small and the coupling $y$ is large, the presence of the extra scalars $\varphi_i$ gives sizeable one-loop vertex corrections to the $Z\mu\mu$ coupling. Such a loop effect modifies the ratio $R_{\mu e}$ in a significant way for large enough values of the coupling $y$. This implies that EWPT constraints push the viable parameter space into the large mixing region, meaning large values of $\xi$ to allow small values of $y$, except in two limiting regions: when $m_\phi$ is small (hence $y$ is also small), which suppresses the loop effect, and when $m_\phi\sim1750$ GeV where there is a dip due to an accidental cancellation between the different loop contributions to $R_{\mu e}$.
    
   \item \textbf{Perturbative unitarity in $\kappa$:} For a given $m_S$, increasing the mixing $\xi$ implies decreasing the mass of the lightest eigenstate $\varphi_1$ while increasing $\kappa$. When a trilinear coupling is large compared to the mass scales in the theory, it can be constrained by perturbative unitarity. Our analysis is similar to the one in \cite{Capdevilla:2021rwo,Capdevilla:2020qel}. We consider the scattering of the lightest eigenstate $\varphi_1 \varphi_1 \to \varphi_1 \varphi_1$ via Higgs exchange. From the scattering amplitude we calculate the partial wave expansion coefficient
  \be
  a_0=\frac{1}{32\pi}\sqrt{\frac{4 p k}{s}} \int d \cos \theta \, \mathcal{M}(\cos \theta),
  \ee
 where $p,k,s$ are the initial momentum, final momentum, and center of mass energy of the process, and $\theta$ is the scattering angle. All these quantities are defined in the center of mass frame. To find the constraints from unitarity on $\kappa$, for a given set of masses $m_S$ and $m_\phi$, we find the $\xi$ value that saturates the unitarity condition $|{\rm Re}(a_0)| < 1/2$.
 
 \item \textbf{LHC:} Due to the four new scalars $\varphi_i$ one gets contributions to $3+4\mu$ production in proton collisions. 
  This happens through radiating some of the scalars $\varphi_i$ off muon lines in charged and neutral Drell-Yan production (through similar diagrams to those in Fig.~\ref{f.lhc}) or by pair producing scalars that subsequently decay down to muons and/or neutrinos. 
  Similar to the singlet-strahlung production analysis in \sref{four-mu}, we derive the projected sensitivity of the LHC and HL-LHC. Unlike the singlet-only analysis we only require that the total BSM contribution to the $3+4 \mu$ production rate is smaller than a $2\sigma$ upward fluctuation of the SM contribution. This conservative choice reflects the fact that the BSM production of 3 and 4 muons includes both resonant and non-resonant processes, and a more comprehensive analysis of this scenario may yield stronger constraints. 
The BSM $3+4\mu$ production cross section is dominated by the charged component of the doublet $\Phi$, which is why the constraint vanishes for large $m_\Phi$.
  The zigzag behavior comes from the fact that the neutral scalars contribute and can in principle interfere with the contributions from the charged scalar in non-trivial ways. In any case, the trend holds that the constraints vanish for large $m_\Phi$. 
 If the singlet has significant couplings to particles other than muons, e.g.\ invisible sector states, then these bounds would be weakened.
 
\end{itemize}

\end{appendix}

\bibliographystyle{JHEP}
\bibliography{SingletPaper}

\end{document}